\documentclass[sigconf,nonacm]{acmart}
\setcopyright{none}
\renewcommand\footnotetextcopyrightpermission[1]{}
\usepackage[T1]{fontenc}
\usepackage[utf8]{inputenc}
\usepackage{amsmath}
\usepackage{booktabs,longtable,array,calc,needspace}
\usepackage{graphicx}
\usepackage{microtype}
\usepackage{fvextra}
\usepackage{enumitem}
\usepackage{xurl}
\usepackage{newunicodechar}
\newunicodechar{§}{\S}
\newunicodechar{×}{\ensuremath{\times}}
\newunicodechar{ε}{\ensuremath{\varepsilon}}
\newunicodechar{–}{--}
\newunicodechar{—}{---}
\newunicodechar{≤}{\ensuremath{\le}}
\newunicodechar{°}{\ensuremath{^\circ}}
\newunicodechar{∞}{\ensuremath{\infty}}
\newunicodechar{Δ}{\ensuremath{\Delta}}
\newunicodechar{𝜀}{\ensuremath{\varepsilon}}
\newunicodechar{𝐿}{\ensuremath{L}}
\newunicodechar{𝐸}{\ensuremath{E}}
\newunicodechar{𝐾}{\ensuremath{K}}
\newunicodechar{𝑆}{\ensuremath{S}}
\newunicodechar{𝑥}{\ensuremath{x}}
\newunicodechar{𝑡}{\ensuremath{t}}
\newunicodechar{𝑖}{\ensuremath{i}}
\newunicodechar{𝑁}{\ensuremath{N}}
\newunicodechar{𝑀}{\ensuremath{M}}
\newunicodechar{𝒳}{\ensuremath{\mathcal{X}}}
\newunicodechar{ℝ}{\ensuremath{\mathbb{R}}}
\newunicodechar{𝜏}{\ensuremath{\tau}}
\newunicodechar{𝑗}{\ensuremath{j}}
\newunicodechar{𝑦}{\ensuremath{y}}
\newunicodechar{𝑚}{\ensuremath{m}}
\newunicodechar{𝜈}{\ensuremath{\nu}}
\newunicodechar{𝛿}{\ensuremath{\delta}}
\newunicodechar{Δ}{\ensuremath{\Delta}}

\DefineVerbatimEnvironment{Highlighting}{Verbatim}{breaklines,breakanywhere,fontsize=\footnotesize,commandchars=\\\{\}}
\DefineVerbatimEnvironment{Verbatim}{Verbatim}{breaklines,breakanywhere,fontsize=\footnotesize}
\RecustomVerbatimEnvironment{verbatim}{Verbatim}{breaklines,breakanywhere,fontsize=\scriptsize}
\providecommand{\tightlist}{\setlength{\itemsep}{0pt}\setlength{\parskip}{0pt}}
\setlist{nosep,leftmargin=*}
\AtBeginEnvironment{longtable}{\scriptsize}
\AtBeginEnvironment{tabular}{\scriptsize}
\makeatletter
\def\@titlefont{\Huge\bfseries}
\@ifclasslater{acmart}{2026/05/30}{%
  \ClassInfo{bbsolver}{Compiling with acmart v2.18 or newer}%
}{%
  \ClassWarningNoLine{bbsolver}{This source is formatted for acmart v2.18 or newer; your local TeX distribution has an older acmart. The preview PDF may differ slightly from a current TeX Live/Overleaf build}%
}
\makeatother

\ExplSyntaxOn
\tl_new:N \l__bb_path_tl
\cs_set_eq:NN \__bb_orig_texttt:n \texttt
\RenewDocumentCommand \texttt { m }
  { \__bb_orig_texttt:n { \__bb_pathbreak:n {#1} } }
\cs_new_protected:Npn \__bb_pathbreak:n #1
  {
    \tl_set:Nn \l__bb_path_tl {#1}
    \regex_replace_all:nnN { ([/.\-]) } { \1 \c{allowbreak} } \l__bb_path_tl
    \regex_replace_all:nnN { \c{_} } { \c{_} \c{allowbreak} } \l__bb_path_tl
    \tl_use:N \l__bb_path_tl
  }
\ExplSyntaxOff

\title{bbsolver: A Unified Error-Bounded Spatiotemporal Optimization Solver for Key Timing and Topology-Consistent Vector Paths}
\author{Ilya Gusinski}
\affiliation{\institution{IVG Design}\country{USA}}
\email{ilyav@gusinski.us}
\date{2026-05-29}

\begin{document}

\begin{abstract}
Dense sampling records what an animation system actually evaluated, but it produces a poor final representation: every sampled frame can become a key, edit handles become noisy, and animated vector paths remain hard to adjust. Existing reducers usually treat the two axes separately: animation-curve reducers reduce key timing, while curve and path simplifiers reduce geometry. When applied independently to animated paths, these methods can break point identity across frames, change vertex structure over time, or provide no single error budget that covers both timing and shape.

bbsolver frames the task as tolerance-bounded spatiotemporal reduction. A host application, such as After Effects or Blender, samples temporal and spatial animation into a documented JSON bundle; the standalone solver chooses sparse keys, interpolation metadata, and path representation; and the output is accepted only if replayed samples remain within the requested worst-case error. The same solver core can be used by any application that can export samples and write back returned keys or paths.

In After Effects validation, solved keys written back into AE and re-sampled from AE playback reduce a DUIK humanoid walk cycle from 12,684 samples to 540 keys at \(\varepsilon=1\), a 23.5\(\times\) reduction, and an ant rig from 11,956 samples to 653 keys, an 18.3\(\times\) reduction, with maximum errors below 1 px and 1 degree. A Blender-sampled FBX mocap retarget reaches 214 keys from 13,455 samples at \(\varepsilon=3\); baselines tuned to matched measured accuracy require 4.5\(\times\) to 27.5\(\times\) more scalar key entries. For vector paths, bbsolver supports reduction when vertex identity/order is constant over time and diagnostics for variable-vertex-count streams, including a 6.7\(\times\) After Effects-compatible procedural-path compression and exact transition-timing recovery in a diagnostic case.
\end{abstract}

\maketitle
\section{Introduction}\label{introduction}

Production animation systems often contain expressions, procedural rigs, parent constraints, generated paths, simulations, auto-traced outlines, or externally imported motion. Sampling the evaluated result at many frame times is robust because it records what the animation application actually played. It is also costly: dense sampled output is large, visually cluttered, and hard to edit. A useful sparse conversion needs to preserve the evaluated motion while returning an artist-facing representation with fewer keys, fewer path vertices, and predictable error behavior.

Existing reducers usually split this problem. Animation-curve reducers reduce temporal key density, but they do not simplify vector geometry or reason about path topology. Curve and path simplifiers reduce spatial complexity, but when applied independently per frame they can change vertex count, ordering, or point identity over time. For animated vector paths, the practical problem is therefore not just temporal reduction or spatial reduction, but the coupled task of reducing key timing and path representation while preserving a single sampled-error budget.

\subsection{Terminology and system boundary}\label{terminology-and-system-boundary}

This paper uses \emph{host} to mean the digital content-creation (DCC) or animation application that evaluates and stores animation, such as After Effects or Blender. A \emph{host adapter} or \emph{harness} is the surrounding integration code that samples animation from the host, serializes it for the solver, and writes the solved result back into the host's native editable representation.

In production animation, artists often \emph{bake} procedural, expression-driven, or rigged motion by sampling the evaluated result into explicit keyframes. This paper uses \emph{dense sampling} for that input process and \emph{sparse conversion} for bbsolver's output. A \emph{host round-trip evaluation} means writing solved keys back into the host, replaying or re-sampling the host's own result, and comparing that playback against the original dense samples.

bbsolver is \emph{host-agnostic} in the sense that the optimization engine is not tied to one DCC application, plug-in API, or animation package. The solver consumes a documented JSON input bundle, called a \texttt{SampleBundle}, containing sampled temporal and spatial animation data. It runs as a standalone process and returns an optimized JSON output bundle, called a \texttt{KeyBundle}, containing sparse keys, interpolation metadata, path data, and diagnostics. Porting bbsolver to another DCC therefore requires an extraction/writeback adapter for software that can expose sampled animation values and accept editable keys or path data; it does not require rewriting the solver core.

For vector paths, \emph{topology} refers to the discrete structure of the path over time: vertex count, contour order, winding direction, start vertex, and index-level point correspondence for vertices and handles. Here \emph{correspondence} means persistent point identity across time: vertex \(k\) at one sample is intended to represent the same logical path point, with its handles, as vertex \(k\) at another sample. This is distinct from layer-space position, which is the geometry being fitted; a corresponding vertex may move far, and two nearby vertices may still fail to correspond if their identity or ordering differs. Constant-topology paths preserve this structure across samples; variable-topology paths change vertex count or point identity over time.

\subsection{Contributions}\label{contributions}

The relevant literature is broad, spanning curve simplification, Bezier fitting, animation key reduction, vector-animation representations, and editable roto systems. bbsolver combines established ideas from those areas into a practical sparse-conversion system with the following characteristics:

\begin{enumerate}
\def\labelenumi{\arabic{enumi}.}
\tightlist
\item
  \textbf{Host-agnostic solve contract.} The solver core is independent of a particular DCC application. Any host that can export sampled temporal or spatial animation into the documented JSON input schema and write the returned optimized bundle back as native editable keys or paths can use the same solver. The After Effects and Blender integrations are reference adapters, not solver dependencies.
\item
  \textbf{Validated sparse output.} The solver accepts a candidate only after replaying it against the dense source stream under a property-specific error metric.
\item
  \textbf{Host round-trip evaluation.} The headline After Effects results are not only internal solver residuals; the solved keys are written back into After Effects and re-sampled from After Effects playback.
\item
  \textbf{Unified temporal/spatial path support.} The same contract handles scalar/vector channels, spatial motion paths, constant-topology shape paths, and diagnostic variable-topology path streams, accepting key timing and path representation together under one sampled error budget.
\item
  \textbf{Reproducible artifact package.} The public repository includes raw \texttt{bbsm}, \texttt{bbky}, \texttt{verify}, and \texttt{progress.log} bundles for the paper-cited solves, supplementary CSVs for quantitative claims, and the After Effects benchmark project used for full host round trips.
\end{enumerate}

The claim is intentionally narrow: the paper contributes a complete, open artifact that combines these characteristics in one place. Prior work covers many of the individual ingredients; the practical advantage here is the combination of tolerance-bounded dense-sample reduction, editable sparse output, a host-agnostic external solve contract, host round-trip evaluation, and topology-aware vector-path handling.

\section{Mathematical formulation and metrics}\label{mathematical-formulation-and-metrics}

Let a sampled animated property be

\[
S = \{(t_i, x_i)\}_{i=1}^{N}, \qquad x_i \in \mathcal{X},
\]

where \(t_i\) is time and \(x_i\) is a property value. Depending on the property, \(\mathcal{X}\) may be scalar, \(\mathbb{R}^2\), \(\mathbb{R}^3\), color space, or a flattened path representation (the \texttt{shape\_flat} encoding detailed in §3.2). A candidate sparse output is a key sequence

\[
K = \{(\tau_j, y_j, m_j)\}_{j=1}^{M}, \qquad M \ll N,
\]

where \(m_j\) contains interpolation metadata such as hold/linear/Bezier mode, temporal ease, spatial tangents, or path handles. Let \(H_K(t)\) denote the evaluator used to replay \(K\). In an ideal host-verified test, \(H_K\) is the host itself after writeback; in a solver-internal test, it is bbsolver's implementation of the target interpolation model.

For a tolerance \(\varepsilon\), the core acceptance condition is

\[
E_\infty(K; S) = \max_i d\bigl(H_K(t_i), x_i\bigr) \le \varepsilon.
\]

For scalar channels, \(d(a,b)=|a-b|\). For vector channels, the paper reports component-space or Euclidean residuals depending on the host fixture and property kind. For path values, \(d\) is a contour distance: each path is densely resampled into a polyline, and distance is approximated by closest-point projection between the source and reconstructed polylines. The production acceptance metric is the \textbf{symmetric} (undirected) Hausdorff distance,

\[
d_P(P,Q) \approx \max\Bigl\{\, \max_{p \in \widehat{P}} \min_{q \in \widehat{Q}} \|p-q\|_2,\;\; \max_{q \in \widehat{Q}} \min_{p \in \widehat{P}} \|p-q\|_2 \,\Bigr\},
\]

i.e.~the larger of the two directed distances, so that neither an unmatched source point nor an unmatched reconstructed point can pass the budget undetected. Every \texttt{shape\_flat} path residual reported in this paper uses this symmetric form; a single directed distance is used only in internal diagnostics where explicitly noted. The solver can also report an RMS residual, though this paper reports worst-case \(L_\infty\) error throughout:

\[
E_2(K;S) = \sqrt{\frac{1}{N}\sum_i d\bigl(H_K(t_i),x_i\bigr)^2}.
\]

For topology diagnostics, let \(\nu_i\) be the source vertex count at sample \(t_i\), and let \(\hat{\nu}_i\) be the count implied by the reconstructed candidate. Topology transition frames are

\[
T(S) = \{ i : \nu_i \ne \nu_{i-1} \}.
\]

Precision and recall are computed by matching source and output transitions within a frame tolerance \(\delta\). The CS2 transition audit in §5.6 sets \(\delta=2\), but it reports a 0-frame maximum transition error, so its precision/recall result is independent of \(\delta\). This precision/recall diagnostic is meaningful for variable-topology sources, but it does not describe AE production writeback: AE-native path keys require representable index-level point correspondence, so production writeback often projects a variable-topology source onto a uniform-topology output.

The optimization objective is constrained (lexicographic over the cost terms below) rather than a closed-form expression to minimize:

\[
\min_K \; C(K) \quad \text{subject to} \quad E_\infty(K;S) \le \varepsilon,
\]

where \(C(K)\) may include key count, path vertex count, output float volume, user constraints, mandatory keyed frames, and fallback penalties. bbsolver searches for candidate key sets and rejects any output that fails validation. The resulting acceptance condition is a finite-sample guarantee, not a theorem about all continuous times between samples.

\subsection{Conformance proposition}\label{conformance-proposition}

\textbf{Proposition.} Suppose the solver accepts a candidate \(K\) only after evaluating \(H_K(t_i)\) for every validation sample \((t_i,x_i)\in S\) and checking that \(d(H_K(t_i),x_i)\le \varepsilon\). Then the accepted candidate satisfies \(E_\infty(K;S)\le\varepsilon\) on the validation sample set.

\textbf{Proof.} By construction, acceptance requires the inequality \(d(H_K(t_i),x_i)\le\varepsilon\) for each sample index \(i\). Taking the maximum over all validation samples gives \(\max_i d(H_K(t_i),x_i)\le\varepsilon\), which is exactly \(E_\infty(K;S)\le\varepsilon\). This establishes discrete-sample conformance for the evaluator used in the check. It does not guarantee that the continuous-time curve between validation samples stays within \(\varepsilon\), nor that a different host evaluator will replay the same metadata identically. In particular, when \(H_K\) is a different host evaluator --- e.g.~After Effects' path densifier --- the replayed error can exceed \(\varepsilon\), as the noodle path shows in §5.7 (\texttt{ae\_roundtrip\_max\_err\ =\ 1.160} px at \(\varepsilon=1\)); the guarantee is therefore against the in-loop solver evaluator, and the AE round-trip is reported separately as an empirical, not a guaranteed, bound.

\section{Solver architecture}\label{solver-architecture}

The solver is organized as a candidate-fit-validate pipeline rather than as a single simplification primitive. The following outline is implementation-oriented because the contribution is a system contract, not a single algorithm.

\begin{verbatim}
input: SampleBundle S, tolerance eps, property metadata, constraints
output: KeyBundle K or diagnostic fallback

1. Parse and classify each property stream.
   - scalar/vector, spatial motion, color, shape_flat path
   - stable topology, variable topology, unsupported topology change
   - mandatory frames, hold sections, keyed constraints, sharp features

2. Seed mandatory keys.
   - first and last samples
   - discontinuities or holds
   - user-preserved frames
   - path topology or sharp-feature landmarks where applicable

3. Generate candidate key sets.
   - dynamic-programming or split-merge temporal placement
   - local refinement around high residual intervals
   - optional spatial/path vertex candidates for shape streams

4. Fit interpolation metadata.
   - host-compatible linear/Bezier/hold parameters
   - spatial tangents and temporal ease where supported
   - path vertices and handles for shape_flat streams

5. Validate against dense samples.
   - replay candidate with the intended evaluator
   - compute max residual and optional RMS residual
   - reject candidates outside eps

6. Prune or refine.
   - remove redundant keys/vertices only if validation still passes
   - bridge path regions when temporal coherence would otherwise pop
   - fall back to safer dense output if no sparse candidate passes

7. Emit KeyBundle.
   - sparse keys, interpolation metadata, diagnostics, notes
   - solver version/build fields for reproducibility
\end{verbatim}

The validation step is the architectural difference between a reducer that accepts a fit on a local geometric criterion and a bake solver that treats the global tolerance as a contract. The open-source baseline comparison in §5.3 illustrates why this matters: algorithms can accept locally reasonable curve fits while still violating the global residual budget against the original sampled stream.

\Needspace{8\baselineskip}
\begin{quote}\small
\textbf{Sidebar --- host-agnostic in practice.} In the independent ant-rig solve \texttt{req-1779849705256}, After Effects crashed mid-solve after the adapter had already serialized the \texttt{SampleBundle} and launched \texttt{bbsolver} as a separate OS process. The solver continued running, unaware of the host failure, and reproduced the §5.2 result: 653 keys, 18.3× compression, and 389.5 s elapsed time recorded in \texttt{solve\_time\_ms}. This is the operational consequence of the JSON-file process boundary; a pipe-based or in-process integration would have died with the host.
\end{quote}

\subsection{Operating modes: in-loop acceptance vs post-solve verify}\label{operating-modes-in-loop-acceptance-vs-post-solve-verify}

The \(L_\infty\) guarantee enforced in §5.2 / §5.3 comes from the \textbf{in-loop candidate-rejection gate} (Algorithm step 5 above): every fit candidate is evaluated at all source sample times and rejected if it exceeds \(\varepsilon\). This check is part of the solve and cannot be disabled.

The AE harness layers a separate \textbf{post-solve verification pass} on top, gated on \texttt{bbsolver-test-harness.jsx:3923} (\texttt{verifyRoundTrip}). With this enabled, the harness re-applies the solved keys to the comp, re-evaluates AE's own playback at every source sample, and can emit host-evaluated verify reports; these AE round-trip residuals are the \texttt{ae\_roundtrip\_max\_err} column. With it disabled, the artist accepts or rejects visually. The \texttt{verify.json} / \texttt{last\_verify\_card.txt} files shipped under \texttt{corpus/} are \emph{not} these AE reports: they are canonical CLI regenerations (\texttt{bbsolver\ verify} v1.0.1, §8), and the \texttt{cli\_verify\_max\_err} column reads from them.

On the §5.2 walk-cycle (12,684 samples), the post-solve verify adds roughly 2-3 minutes at tight \(\varepsilon\); this overhead shrinks toward the bare inner-solve cost as \(\varepsilon\) loosens:

\begin{table*}[t]
\centering
\scriptsize
\caption{Post-solve verify-pass overhead on the §5.2 walk-cycle (12,684 samples).}\label{tbl:verify-overhead}
\begin{tabular}{@{}
  >{\raggedright\arraybackslash}p{(\textwidth - 8\tabcolsep) * \real{0.1579}}
  >{\raggedleft\arraybackslash}p{(\textwidth - 8\tabcolsep) * \real{0.2105}}
  >{\raggedleft\arraybackslash}p{(\textwidth - 8\tabcolsep) * \real{0.2105}}
  >{\raggedleft\arraybackslash}p{(\textwidth - 8\tabcolsep) * \real{0.2105}}
  >{\raggedleft\arraybackslash}p{(\textwidth - 8\tabcolsep) * \real{0.2105}}@{}}
\toprule\noalign{}
\begin{minipage}[b]{\linewidth}\raggedright
Run (walk-cycle)
\end{minipage} & \begin{minipage}[b]{\linewidth}\raggedleft
bbsolver solve
\end{minipage} & \begin{minipage}[b]{\linewidth}\raggedleft
wall-clock incl.~AE re-eval + verify
\end{minipage} & \begin{minipage}[b]{\linewidth}\raggedleft
overhead
\end{minipage} & \begin{minipage}[b]{\linewidth}\raggedleft
ratio
\end{minipage} \\
\midrule\noalign{}
ε=0.05 (\texttt{req-1779727765498}) & 18.0 m & 20.7 m & +2.7 m & 1.15× \\
ε=1.0 (\texttt{req-1779735092073}) & 15.9 m & 18.3 m & +2.4 m & 1.15× \\
ε=3.0 (\texttt{req-1779733597712}) & 18.3 m & 18.7 m & +0.4 m & 1.02× \\
\end{tabular}
\end{table*}

As Table~\ref{tbl:verify-overhead} shows, the verify pass adds 3-15 \% wall-clock on this scale (dominated by AE re-evaluation, not by \texttt{bbsolver\ verify}); production solves run with \texttt{verifyRoundTrip=false} to skip it. The \(L_\infty\) bound is unaffected because it lives in the always-on inner gate, not in the outer verify pass. Headline accuracy numbers in this paper were produced with the verify pass on, so every ``max\_err \(\le \varepsilon\) on every property'' claim is empirically anchored.

\subsection{Paths and topology}\label{paths-and-topology}

For path streams, bbsolver does not treat temporal key reduction and spatial path simplification as independent post-processes: key timing and the spatial/path representation are accepted or rejected together under the same sampled error budget. This is what the title's ``unified spatiotemporal'' refers to.

A \texttt{shape\_flat} sample encodes a closed flag, a vertex count, vertices, and in/out tangent data as a flat numeric vector. As in §1.1, topology here means discrete path structure over time, especially vertex count and index-level point correspondence. It does not mean layer-space position: position is the geometry being approximated, while correspondence is the identity/order relationship that lets a host interpolate path keys without swapping logical points. For stable-topology paths, the source vertex count and point identity are constant, and bbsolver can jointly reduce temporal keys and path handles. For variable-topology sources, there are two distinct outputs:

\begin{itemize}
\tightlist
\item
  \textbf{Production host-compatible output.} The solver emits a representable path stream for the target host. In AE this means uniform topology across keyed path values, because AE cannot faithfully replay vertex birth/death (vertices appearing or disappearing over time) as ordinary path keys.
\item
  \textbf{Diagnostic per-region output.} The solver may emit keys whose vertex-count trajectory follows the source topology. This is useful for measuring whether topology events are detected and segmented correctly, even when the output is not the production AE writeback path.
\end{itemize}

This distinction prevents overclaiming. Variable topology is not presented as the common case for hand-keyed AE paths. It is a stress case that arises in procedural expressions, auto-trace/vectorization, generated contours, drawing-substitution morphs, geometry caches, and host imports.

\section{Evaluation protocol}\label{evaluation-protocol}

The evaluation uses four evidence tracks.

\textbf{AE round-trip rigs.} The harness samples AE properties, calls bbsolver, writes the resulting keys back into AE, and re-samples AE playback at the original validation times. This measures the real host representation rather than only an internal curve metric.

\textbf{Cross-host FBX validation.} A Blender Python harness samples a retargeted FBX action, sends transform channels through bbsolver, and compares against Blender F-Curve Decimate plus standalone Python ports of two Maya-oriented open-source reducers. This tests whether the solver boundary is actually useful beyond AE. \emph{FBX-roundtrip caveat:} Blender 4.5's stock \texttt{io\_scene\_fbx} importer hardcodes \texttt{LINEAR\_INTERPOLATION\_VALUE} for every imported keyframe (\texttt{import\_fbx.py:715-716,\ 878}). Its exporter likewise writes only baked linear keys (\texttt{export\_fbx\_bin.py:2078}, \texttt{bake\_anim=True}). The Bezier tangent metadata bbsolver emits is therefore \emph{expected} to round-trip through Maya / MotionBuilder / the Autodesk FBX SDK, which preserve per-key tangents, but is collapsed to linear by Blender's own importer; this Bezier-FBX round-trip is by design and is not independently tested in this paper (see §7.1). The cross-host \emph{solver-input} comparison in §5.3 is unaffected, because that path samples Blender's pose-bone \texttt{matrix\_basis} directly rather than through FBX re-import.

\textbf{Static path and vector-path tests.} SVG and \texttt{shape\_flat} fixtures test vertex reduction, path fidelity, and path representability. Some path tests are production writeback cases; others are diagnostic topology cases.

\textbf{Reproducibility corpus.} The public benchmark folder includes one supplementary CSV per quantitative result, a paper corpus of raw solver input/output bundles, external baseline runners, figure-generation scripts, and the AE benchmark project. The AE project requires a licensed AE installation for full host round-trip reproduction, but the serialized \texttt{SampleBundle} corpus allows solver-only reproduction without AE.

\section{Results}\label{results}

\subsection{Production corpus}\label{production-corpus}

The production corpus is not the main reproducibility evidence because some original production histories are private. It is included to show that the solver is not tuned only for synthetic examples. The aggregate is \textbf{publicly auditable} from the shipped \texttt{supplementary/production\_corpus\_per\_run.csv} (one row per solve) and \texttt{supplementary/production\_corpus\_summary.csv} (aggregate row); both ship in the repo. Full regeneration from the raw \texttt{live\_runs/} folders requires the original private development corpora, which are not redistributed; the paper-cited subset is redistributed under \texttt{corpus/}. \texttt{scripts/generate\_production\_corpus\_summary.py} reproduces the aggregate when pointed at a live-runs root via \texttt{\_paths.py} env / CLI overrides.

The deduplicated corpus contains \textbf{203 unique runs} spanning solver versions from \texttt{bbsolver\ 0.1.0} (development snapshots, May 19-22) through \texttt{bbsolver\ 1.0.0} (paper-cited runs, May 25-26). All 203 have \texttt{solve\_time\_ms} recorded; Table~\ref{tbl:production} summarizes the corpus.

\begin{table*}[t]
\centering
\scriptsize
\caption{Deduplicated production corpus (203 runs): reduction-ratio and solve-time statistics.}\label{tbl:production}
\begin{tabular}{@{}
  >{\raggedright\arraybackslash}p{(\textwidth - 2\tabcolsep) * \real{0.4286}}
  >{\raggedleft\arraybackslash}p{(\textwidth - 2\tabcolsep) * \real{0.5714}}@{}}
\toprule\noalign{}
\begin{minipage}[b]{\linewidth}\raggedright
Statistic
\end{minipage} & \begin{minipage}[b]{\linewidth}\raggedleft
Value
\end{minipage} \\
\midrule\noalign{}
Total runs & 203 \\
Runs touching a path property (Vector Shape / Mask Path / \texttt{shape\_flat}) & 123 (60.6\%) \\
Runs with topology / replacement diagnostic notes & 9 (4.4\%) \\
Median input samples / run & 430 \\
Median output keys / run & 57 \\
\textbf{Median sample-to-key reduction ratio} & \textbf{5.51×} \\
Mean reduction ratio & 32.79× \\
10th-percentile reduction & 1.99× \\
90th-percentile reduction & 120.2× \\
Median solve time & 824 ms \\
90th-percentile solve time & 68.4 s \\
Maximum solve time observed & 1,096.2 s \\
Sum input samples / sum output keys (overall) & 174,350 / 22,318 = 7.81× \\
\end{tabular}
\end{table*}

Path-aware behavior is a recurring production concern rather than a single synthetic case: 60.6\% of runs touch a path property, and 4.4\% carry explicit replacement / canonical-path / per-region topology notes. The long solve-time tail (max \textasciitilde18 minutes) is dominated by path-heavy variable-topology bakes; the median remains under one second. The mean reduction ratio (32.79×) is skewed upward by near-static properties that collapse to two keys, which is why the median (5.51×) is the headline figure.

\begin{figure*}[t]
\centering
\includegraphics[width=0.95\textwidth]{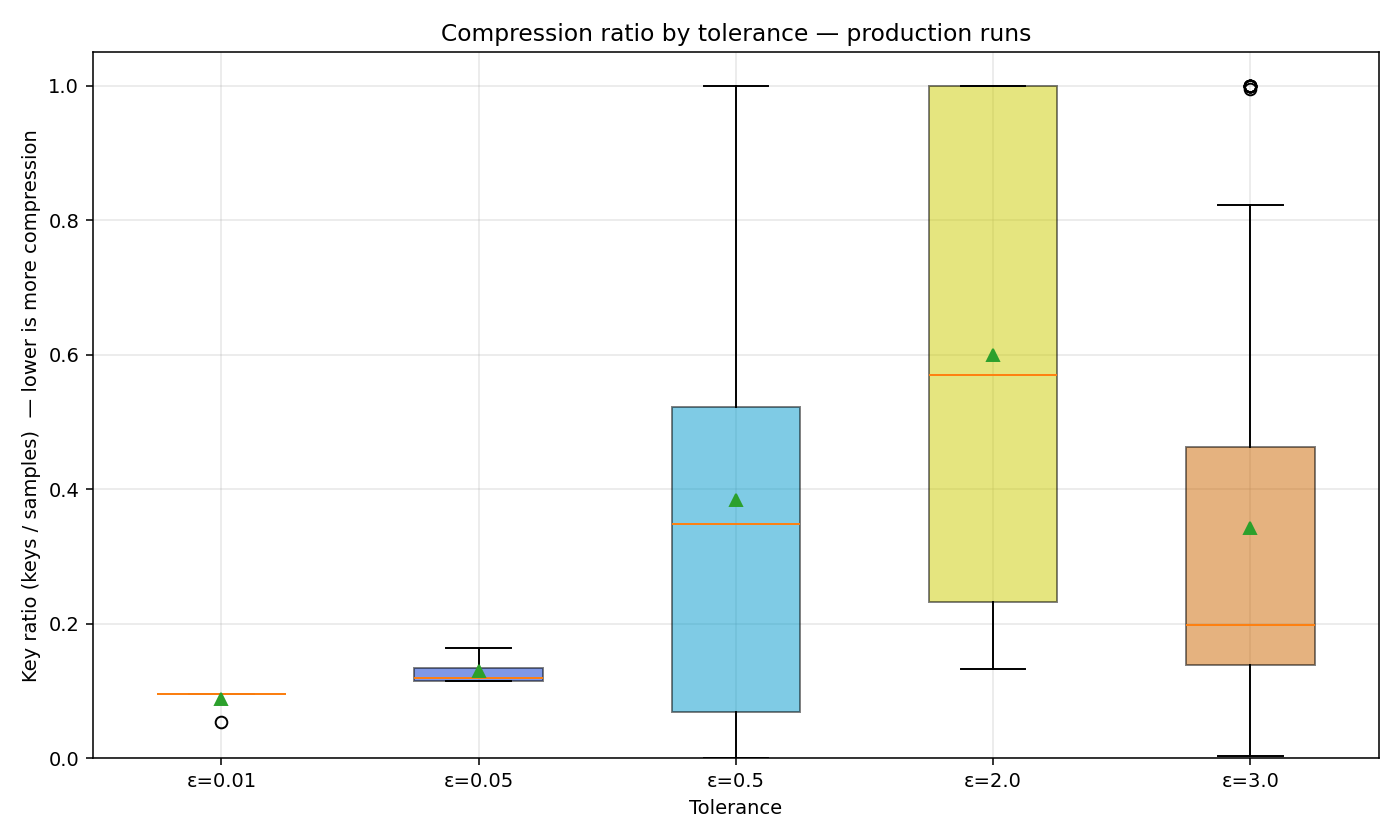}
\caption{Production compression by tolerance.}\label{fig:prod-compression}
\end{figure*}

This corpus is a mixed-version \emph{convenience sample}, not a fixed-budget fidelity benchmark. Roughly 91\% of the 203 runs come from development snapshots (\texttt{bbsolver\ 0.1.0}), the per-run requested \(\varepsilon\) is not recorded in the public CSV, and a minority of runs solve at loose or unspecified tolerances that exceed a few pixels of residual. The reduction ratios here therefore describe operational throughput across heterogeneous tolerances; the controlled accuracy-at-fixed-\(\varepsilon\) claims live in §5.2--§5.3, and the median 5.51× should not be read as a quality result at a single budget. Fig.~\ref{fig:prod-compression} plots achieved compression against tolerance across the corpus.

\subsection{After Effects round-trip: humanoid and ant rigs}\label{after-effects-round-trip-humanoid-and-ant-rigs}

The strongest common-case evidence comes from fixed-topology AE round-trip validation. Two rig fixtures were sampled, solved, written back into AE, and verified by re-evaluating AE playback; Table~\ref{tbl:ae-roundtrip} reports the results.

\begin{table*}[t]
\centering
\scriptsize
\caption{After Effects round-trip results for the DUIK humanoid and ant rigs.}\label{tbl:ae-roundtrip}
\begin{tabular}{@{}
  >{\raggedright\arraybackslash}p{(\textwidth - 10\tabcolsep) * \real{0.1304}}
  >{\raggedleft\arraybackslash}p{(\textwidth - 10\tabcolsep) * \real{0.1739}}
  >{\raggedleft\arraybackslash}p{(\textwidth - 10\tabcolsep) * \real{0.1739}}
  >{\raggedleft\arraybackslash}p{(\textwidth - 10\tabcolsep) * \real{0.1739}}
  >{\raggedleft\arraybackslash}p{(\textwidth - 10\tabcolsep) * \real{0.1739}}
  >{\raggedleft\arraybackslash}p{(\textwidth - 10\tabcolsep) * \real{0.1739}}@{}}
\toprule\noalign{}
\begin{minipage}[b]{\linewidth}\raggedright
Fixture
\end{minipage} & \begin{minipage}[b]{\linewidth}\raggedleft
Source samples
\end{minipage} & \begin{minipage}[b]{\linewidth}\raggedleft
Output keys
\end{minipage} & \begin{minipage}[b]{\linewidth}\raggedleft
Reduction
\end{minipage} & \begin{minipage}[b]{\linewidth}\raggedleft
Max Position error
\end{minipage} & \begin{minipage}[b]{\linewidth}\raggedleft
Max Rotation error
\end{minipage} \\
\midrule\noalign{}
DUIK humanoid, \(\varepsilon=1\) & 12,684 & 540 & 23.5× & 0.978 px & 0.978 deg \\
DUIK humanoid, \(\varepsilon=3\) & 12,684 & 382 & 33.2× & 2.881 px & 2.825 deg \\
Ant rig, \(\varepsilon=1\) & 11,956 & 653 & 18.3× & 0.987 px & 0.986 deg \\
\end{tabular}
\end{table*}

\begin{figure*}[t]
\centering
\includegraphics[width=0.95\textwidth]{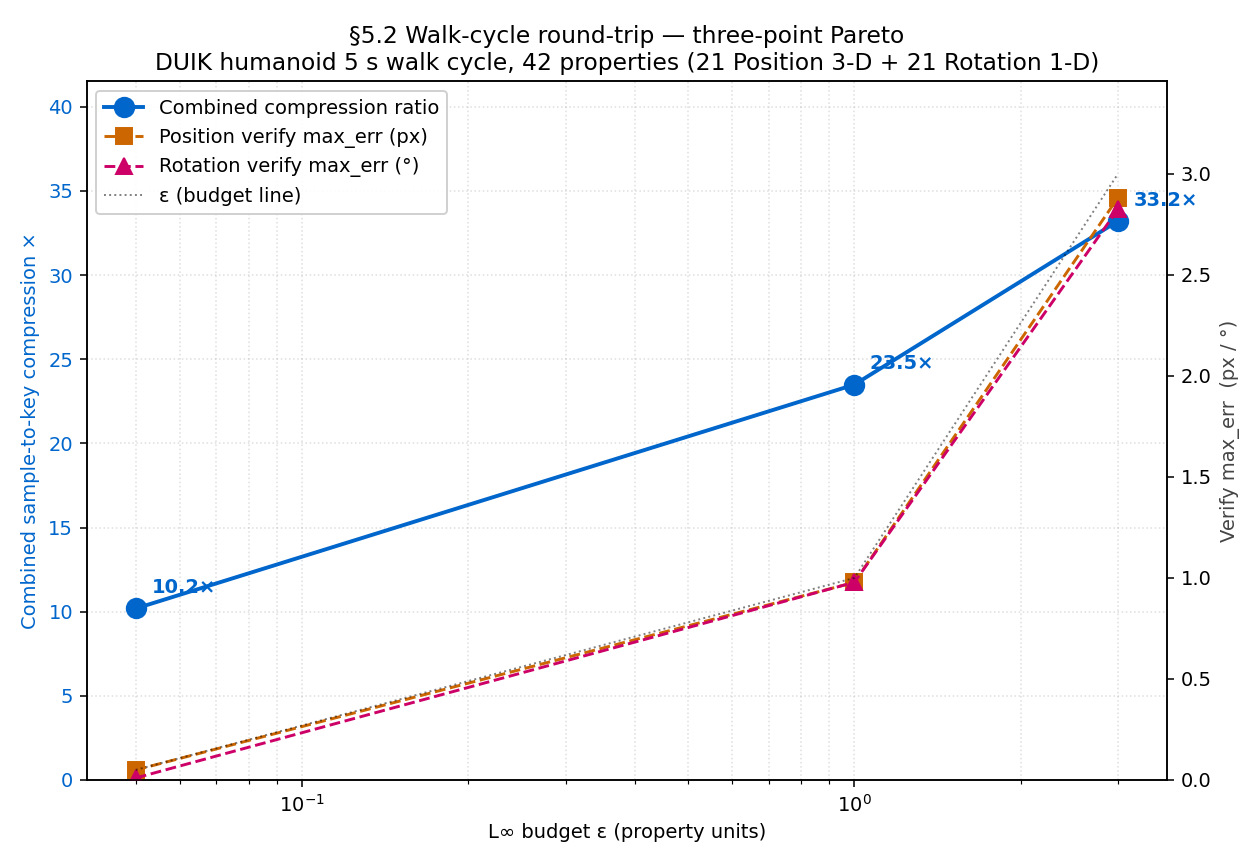}
\caption{DUIK humanoid three-point tolerance sweep.}\label{fig:walk-pareto}
\end{figure*}

\begin{figure*}[t]
\centering
\includegraphics[width=0.95\textwidth]{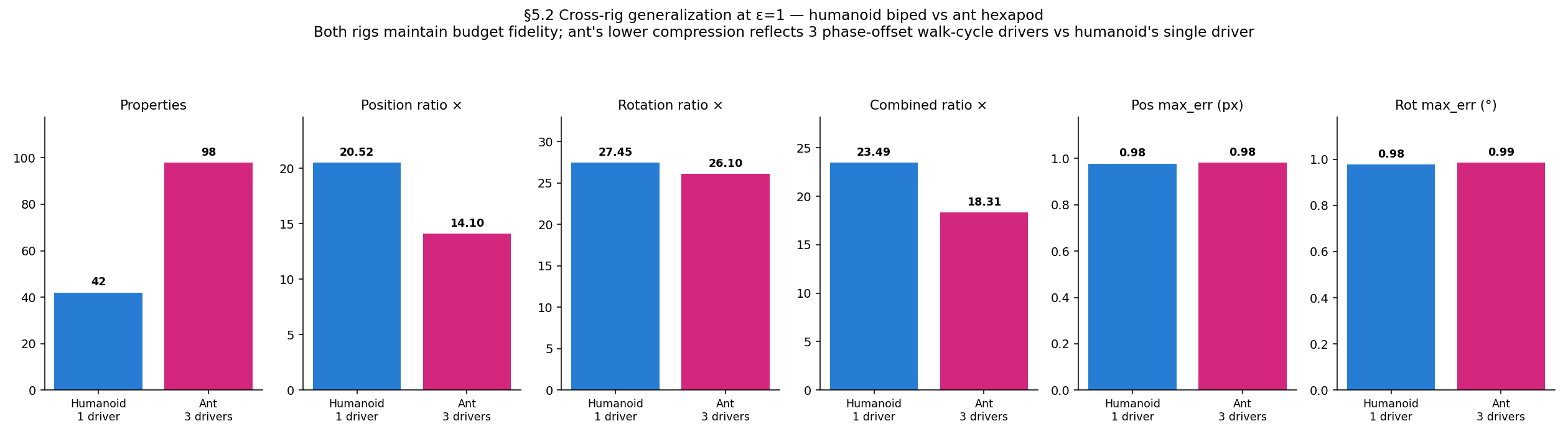}
\caption{Humanoid and ant round-trip comparison.}\label{fig:humanoid-ant}
\end{figure*}

At the AE harness's tighter default rotation tolerance, the DUIK fixture produces 1,245 keys, a 10.2× reduction, with max position error 0.052 px and max rotation error 0.010 deg. This tighter-tolerance result is a useful conservative baseline; the \(\varepsilon=1\) row in the table above is the primary production-tolerance result. (In the per-property supplementary CSVs the boolean \texttt{ok} column is evaluated against the harness's tight \emph{default} rotation tolerance rather than the headline ε, so a property can read \texttt{ok=false} at the default while still satisfying the looser headline ε; read \texttt{ok} together with the ε of its row.) Fig.~\ref{fig:walk-pareto} shows the DUIK three-point tolerance sweep, and Fig.~\ref{fig:humanoid-ant} the humanoid-versus-ant round-trip comparison.

\subsection{Cross-host validation: FBX mocap retarget via Blender}\label{cross-host-validation-fbx-mocap-retarget-via-blender}

A Blender harness sampled a 45-property FBX mocap retarget fixture at 299 samples per property, for 13,455 source samples total. The comparison includes bbsolver, Blender F-Curve Decimate, a standalone Python port of \texttt{robertjoosten/maya-keyframe-reduction}'s Paper.js/Schneider-style reducer, and a standalone Python port of Toolchefs \texttt{keyReducer}. The raw input \texttt{SampleBundle} (\texttt{retarget\_full\_size.bbsm.json}, 45 properties × 299 samples = 13,455 source samples) ships under \texttt{fbx\_mocap\_retarget\_full\_size/pose\_sampled\_blender\_action/}, and \texttt{scripts/\_paths.py} resolves it automatically from a clone. It is a dense sampling of a Mixamo-derived retargeted FBX action; the underlying Mixamo animation is governed by Adobe's Mixamo / Creative Cloud terms, and what ships here is the derived \texttt{SampleBundle}. The bbsolver, Joosten, and Toolchefs branches of the §5.3 comparison therefore re-solve end to end from the artifact with no external input; only the Blender F-Curve Decimate baseline additionally needs a Blender install to re-run its \texttt{bpy.ops.graph.decimate} sweep. The per-method, per-tolerance solver outputs that back every number below also ship under \texttt{fbx\_mocap\_retarget\_full\_size/sweep\_outputs/}, and the \texttt{fbx\_mocap\_*.csv} files in \texttt{supplementary/} summarize them, so the published comparison is auditable both from the shipped outputs and by re-solving from the shipped raw input. All four methods treat rotation as independent per-axis Euler channels (errors reported per channel, in degrees) and position as per-axis channels (in pixels); none applies a quaternion-aware reduction, so the degree-space \(L_\infty\) comparison is apples-to-apples.

\begin{figure*}[t]
\centering
\includegraphics[width=0.95\textwidth]{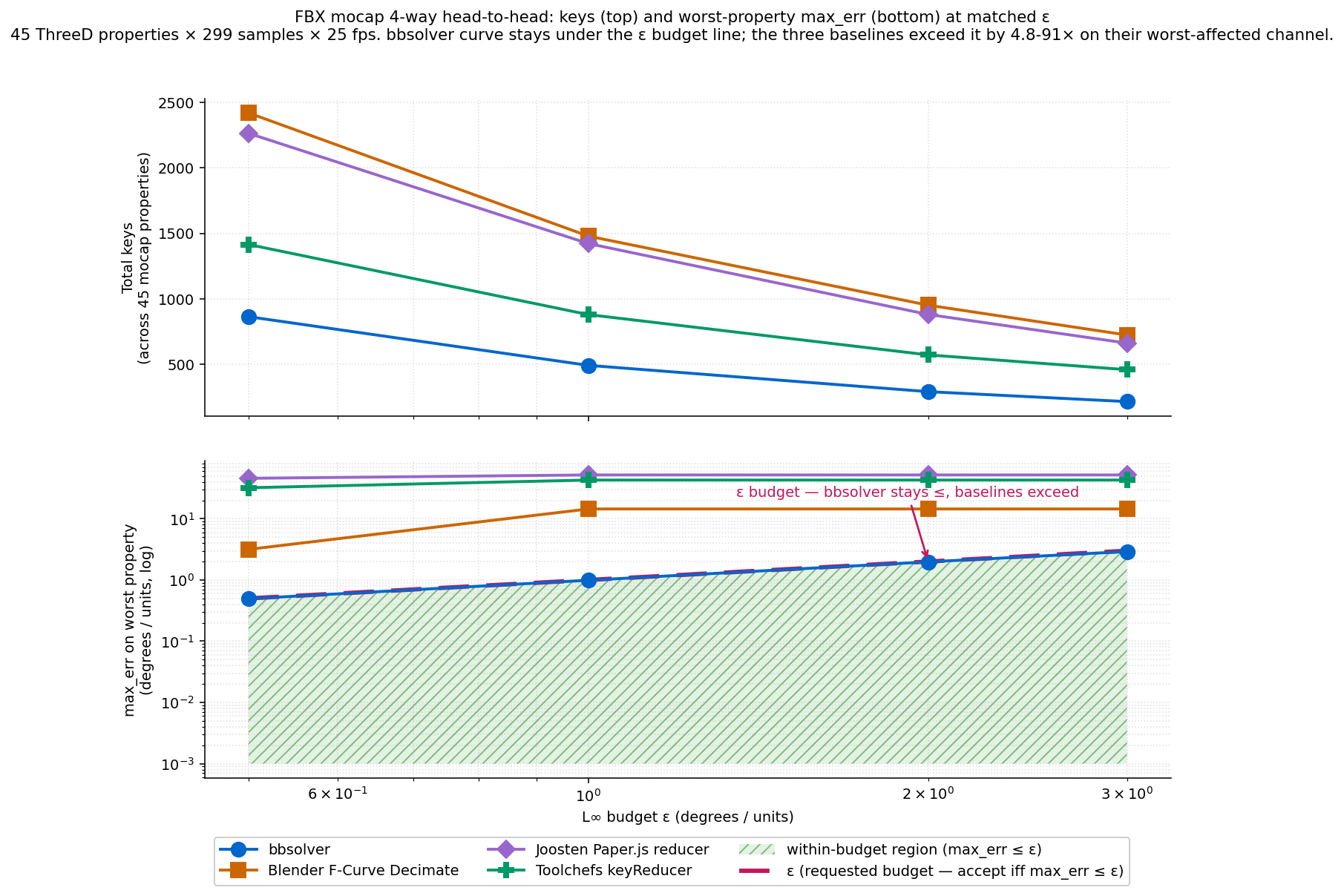}
\caption{FBX mocap Pareto comparison.}\label{fig:fbx-pareto}
\end{figure*}

At matched requested \(\varepsilon\), bbsolver is the only tested method that stays inside budget on every property (Table~\ref{tbl:fbx-matched}; Fig.~\ref{fig:fbx-pareto}).

\begin{table*}[t]
\centering
\scriptsize
\caption{Cross-host FBX comparison at matched requested \(\varepsilon=1\).}\label{tbl:fbx-matched}
\begin{tabular}{@{}
  >{\raggedright\arraybackslash}p{(\textwidth - 4\tabcolsep) * \real{0.2727}}
  >{\raggedleft\arraybackslash}p{(\textwidth - 4\tabcolsep) * \real{0.3636}}
  >{\raggedleft\arraybackslash}p{(\textwidth - 4\tabcolsep) * \real{0.3636}}@{}}
\toprule\noalign{}
\begin{minipage}[b]{\linewidth}\raggedright
Method at \(\varepsilon=1\)
\end{minipage} & \begin{minipage}[b]{\linewidth}\raggedleft
Output keys (vector / scalar-eq)
\end{minipage} & \begin{minipage}[b]{\linewidth}\raggedleft
Worst max error
\end{minipage} \\
\midrule\noalign{}
bbsolver & 491 / 1,473 & 0.990 \\
Blender Decimate & --- / 1,479 & 14.440 \\
Joosten reducer & --- / 1,422 & 51.930 \\
Toolchefs reducer & --- / 879 & 42.830 \\
\end{tabular}
\end{table*}

\textbf{Key-count metric note.} bbsolver emits one shared-timing key per ThreeD (three-dimensional) property (an animator edit point that controls all three dimensions at the same instant); the open-source baselines emit independent scalar F-curve keys per dimension. The ``scalar-equivalent'' column above counts each bbsolver ThreeD key three times to match the baselines' per-axis bookkeeping, giving an apples-to-apples \emph{storage / insertion-count} comparison. On the \textbf{vector} metric (animator edit points), bbsolver's 491 vector keys expand to \textasciitilde1,473 scalar entries --- essentially identical to Blender's 1,479 at the same \(\varepsilon\). The substantive headline at matched \(\varepsilon\) is therefore \textbf{accuracy}, not key count: bbsolver achieves a worst max error of \textasciitilde0.99 (within the \(\varepsilon=1\) budget) where the open-source methods exceed budget by 14×--52×.

\begin{figure*}[t]
\centering
\includegraphics[width=0.95\textwidth]{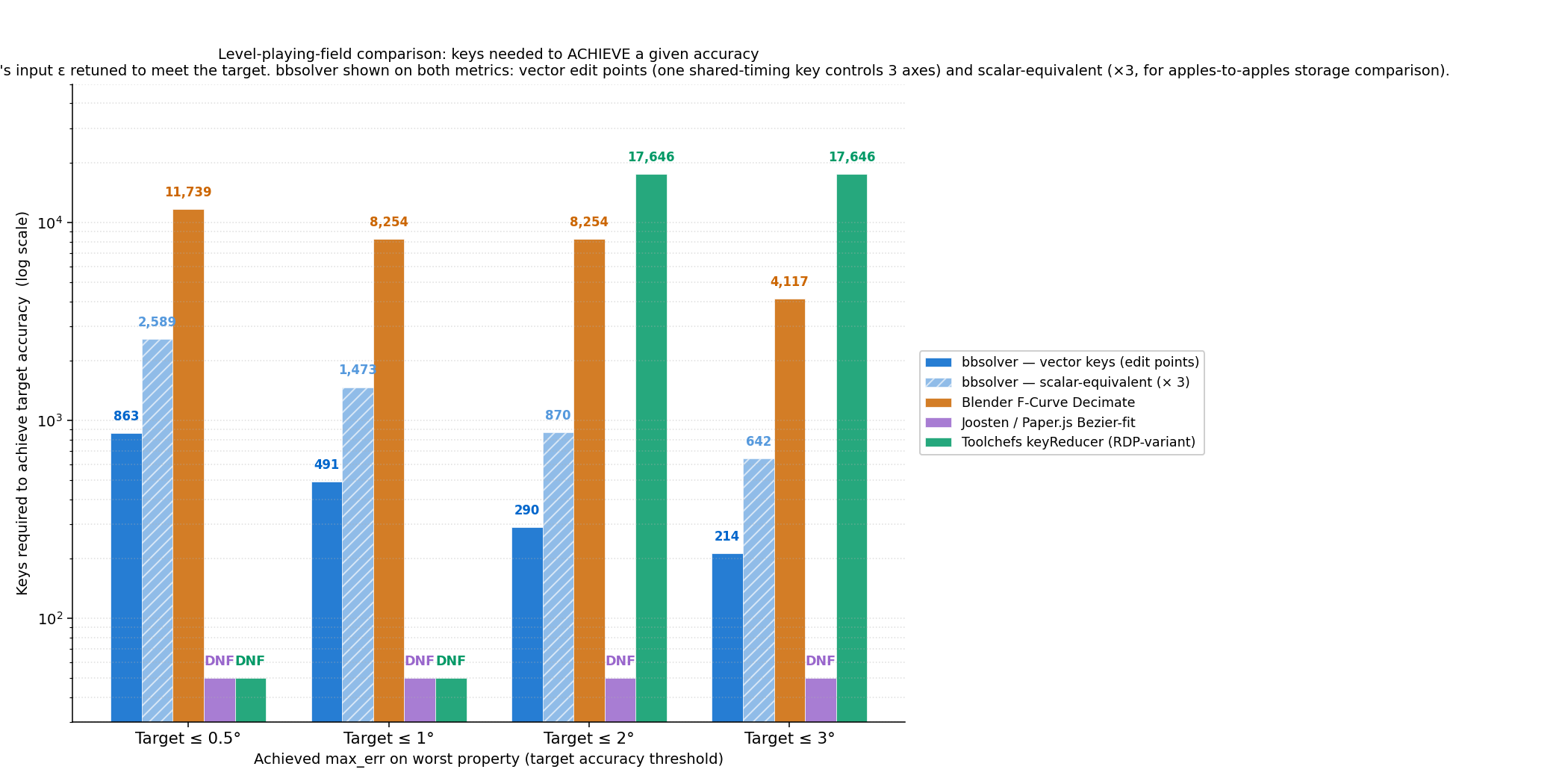}
\caption{Level-playing-field comparison: key counts required to achieve each measured accuracy target.}\label{fig:fbx-lpf}
\end{figure*}

A second comparison (Table~\ref{tbl:fbx-lpf}, Fig.~\ref{fig:fbx-lpf}) retunes each baseline's input tolerance to achieve the requested measured fidelity and then compares key counts on both metrics:

\begin{table*}[t]
\centering
\scriptsize
\caption{Level-playing-field comparison: key counts required to reach each measured-accuracy target.}\label{tbl:fbx-lpf}
\begin{tabular}{@{}
  >{\raggedleft\arraybackslash}p{(\textwidth - 8\tabcolsep) * \real{0.2000}}
  >{\raggedleft\arraybackslash}p{(\textwidth - 8\tabcolsep) * \real{0.2000}}
  >{\raggedleft\arraybackslash}p{(\textwidth - 8\tabcolsep) * \real{0.2000}}
  >{\raggedleft\arraybackslash}p{(\textwidth - 8\tabcolsep) * \real{0.2000}}
  >{\raggedleft\arraybackslash}p{(\textwidth - 8\tabcolsep) * \real{0.2000}}@{}}
\toprule\noalign{}
\begin{minipage}[b]{\linewidth}\raggedleft
Target max\_err
\end{minipage} & \begin{minipage}[b]{\linewidth}\raggedleft
bbsolver (vector / scalar-eq)
\end{minipage} & \begin{minipage}[b]{\linewidth}\raggedleft
Blender (vec× / scalar-eq×)
\end{minipage} & \begin{minipage}[b]{\linewidth}\raggedleft
Joosten
\end{minipage} & \begin{minipage}[b]{\linewidth}\raggedleft
Toolchefs
\end{minipage} \\
\midrule\noalign{}
\(\le\) 0.5° & \textbf{863 / 2,589} & 11,739 (13.6× / \textbf{4.5×}) & DNF & DNF \\
\(\le\) 1.0° & \textbf{491 / 1,473} & 8,254 (16.8× / \textbf{5.6×}) & DNF & DNF \\
\(\le\) 2.0° & \textbf{290 / 870} & 8,254 (28.5× / \textbf{9.5×}) & DNF & 17,646 (60.8× / \textbf{20.3×}) \\
\(\le\) 3.0° & \textbf{214 / 642} & 4,117 (19.2× / \textbf{6.4×}) & DNF & 17,646 (82.5× / \textbf{27.5×}) \\
\end{tabular}
\end{table*}

DNF means no input tolerance in the tested 12-point sweep achieved the target; it should be read as an empirical sweep result, not a mathematical impossibility. On the apples-to-apples scalar-equivalent metric, bbsolver uses \textbf{4.5× to 27.5× fewer key entries} than the baselines that can reach each target. On animator edit points the gap is \textbf{13.6× to 82.5×}, but that ratio includes the legitimate but distinct shared-timing-key benefit. Either way the architectural conclusion holds: per-curve geometric criteria need many more samples to meet a global \(L_\infty\) budget than a validated DP that fits with that budget as its acceptance gate.

\textbf{Error-metric-space and sweep-granularity caveats.} Two effects beyond raw algorithmic efficiency contribute to this gap and should be read alongside it. First, bbsolver's acceptance gate measures residuals in value space (an \(L_\infty\) over property values), whereas the two Maya-derived ports make keep/split decisions on Euclidean distance in a mixed (time, value) plane, and Blender Decimate's ratio control is not a value-space budget at all; part of the matched-ε gap and the DNF outcomes therefore reflects this objective mismatch, which the level-playing-field table mitigates by retuning each baseline to a measured value-space target. Second, Blender's input-tolerance sweep is discrete. The \(\le\) 1.0° and \(\le\) 2.0° rows both resolve to the same 8,254-key / 0.747° operating point, because the next coarser sweep point overshoots 2.0°. The Blender ratios at those two targets are therefore computed against an over-tight solve. We report worst-case (\(L_\infty\)) error throughout because that is the quantity the contract bounds; baseline \emph{median} error is lower than its worst case but is not the budgeted quantity.

\subsection{Static SVG path simplification vs Illustrator}\label{static-svg-path-simplification-vs-illustrator}

\begin{figure*}[t]
\centering
\includegraphics[width=0.95\textwidth]{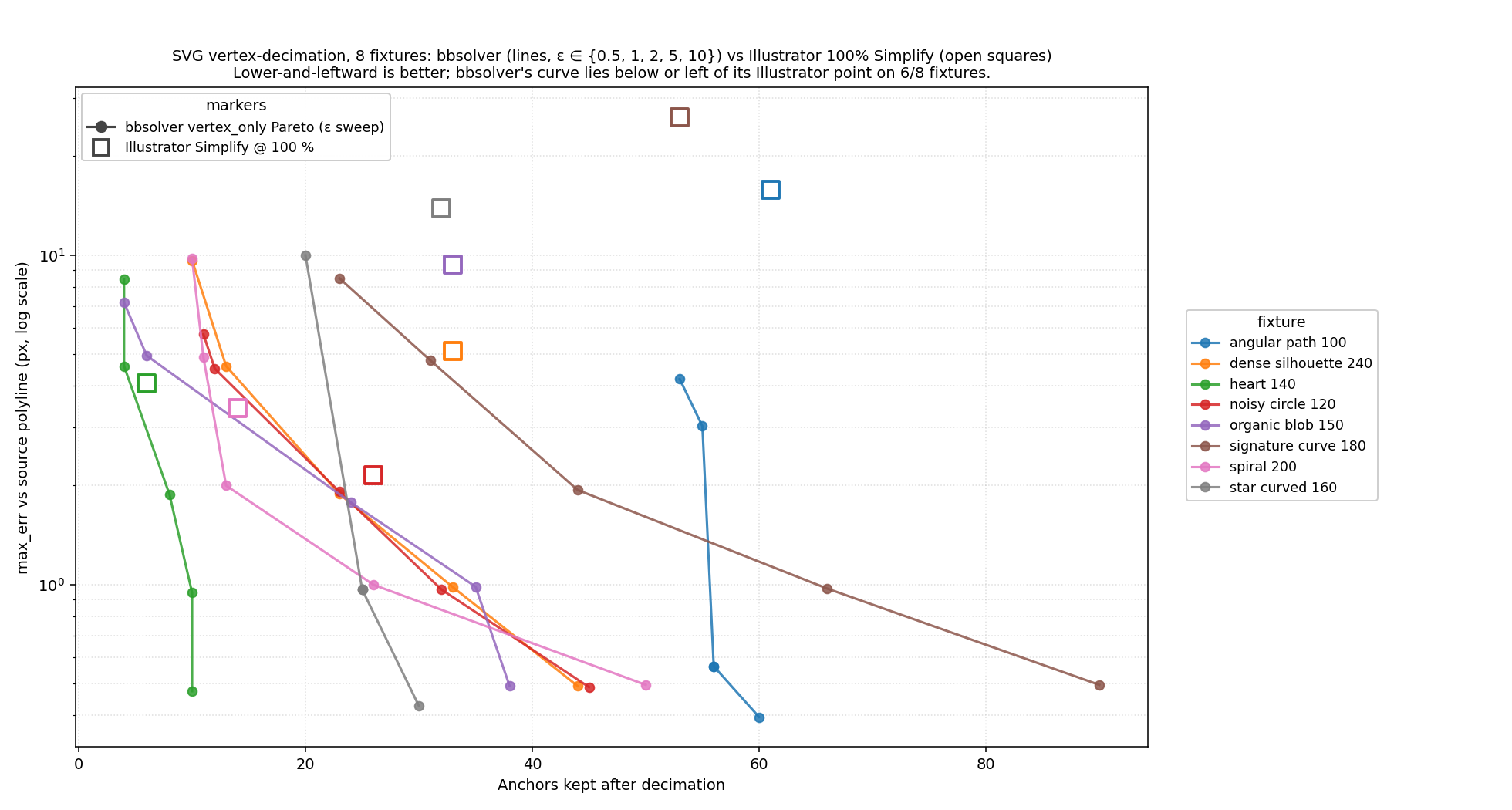}
\caption{Static SVG decimation summary.}\label{fig:svg}
\end{figure*}

Static SVG tests compare bbsolver's vertex-only mode against Adobe Illustrator's \texttt{Object\ \textgreater{}\ Path\ \textgreater{}\ Simplify} workflow. Illustrator is a useful practitioner baseline, but its UI does not expose the same explicit error contract. The result is therefore framed as a practical comparison rather than a controlled head-to-head benchmark; Fig.~\ref{fig:svg} summarizes the eight SVG fixtures.

Representative examples at matched or similar anchor counts:

\begin{itemize}
\tightlist
\item
  \texttt{angular\_path\_100}: Illustrator uses 61 anchors with 15.79 px error; bbsolver uses 60 anchors with 0.40 px error at \(\varepsilon=0.5\).
\item
  \texttt{dense\_silhouette\_240}: Illustrator uses 33 anchors with 5.12 px error; bbsolver uses 33 anchors with 0.98 px error at \(\varepsilon=1\).
\item
  \texttt{star\_curved\_160}: Illustrator uses 32 anchors with 13.84 px error; bbsolver uses 30 anchors with 0.43 px error at \(\varepsilon=0.5\).
\end{itemize}

\textbf{Methodology note.} bbsolver's default config sets \texttt{allow\_path\_spatial\_fit:\ true} and \texttt{path\_preserve\_sharp\_corners:\ true} with a 90° angle threshold. The first is contraindicated for static SVG: it pre-collapses near-collinear vertices before the vertex-only pass runs. That pre-collapse is designed for animation-context inputs, where every vertex is sampled from a continuous trajectory rather than placed deliberately by an artist or auto-trace. The SVG runs in \texttt{corpus/} use \texttt{allow\_path\_spatial\_fit:\ false}. The sharp-corner threshold protects vertices below 90° (e.g.~the five outer tips of \texttt{star\_curved\_160} at \textasciitilde36°) but treats the five inner notches (\textasciitilde108°) as smoothable; widening \texttt{path\_sharp\_corner\_angle\_deg} would preserve both at the cost of over-protecting noisy traced shapes.

\subsection{Constant-topology path reduction: CS1}\label{constant-topology-path-reduction-cs1}

CS1 is a constant-topology animated path. The source has 290 samples and 56 vertices. At \(\varepsilon=1\), bbsolver emits 84 keys and preserves the 56-vertex topology. Per-frame Schneider and Ramer-Douglas-Peucker (RDP) baselines simplify each frame independently, so the vertex count jumps discontinuously between frames (``pops''): the audit records 64 Schneider pops and 62 RDP pops.

\begin{figure*}[t]
\centering
\includegraphics[width=0.95\textwidth]{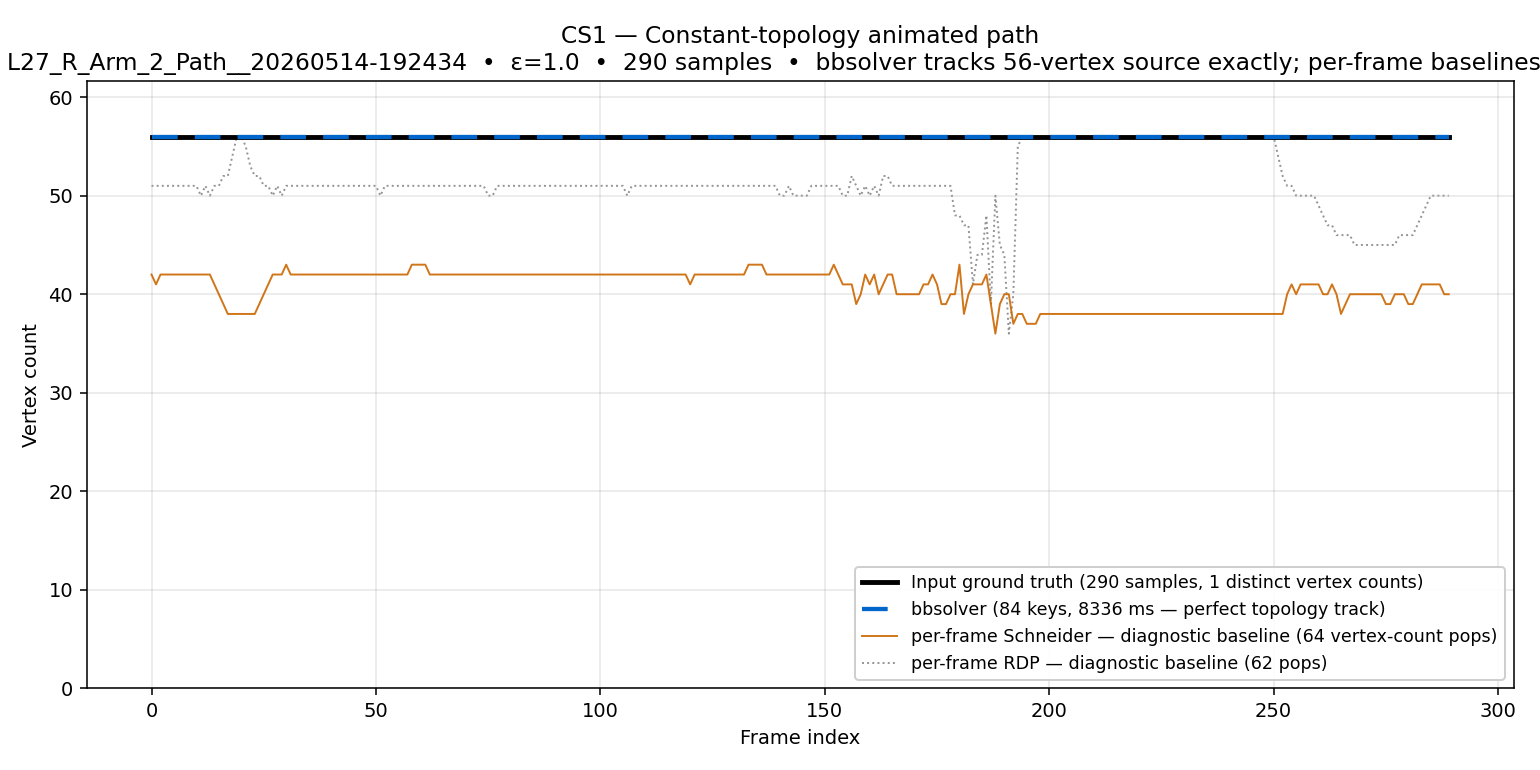}
\caption{CS1 constant-topology path comparison.}\label{fig:cs1}
\end{figure*}

This case (Fig.~\ref{fig:cs1}) supports temporal path reduction and topology stability. It does not claim vertex-count reduction in CS1; that is evaluated separately in oversampled and static SVG path tests.

\subsection{Variable-topology diagnostic: CS2}\label{variable-topology-diagnostic-cs2}

CS2 is a diagnostic stress case for variable-topology input. At \(\varepsilon=1\), the per-region diagnostic solve emits 44 keys and preserves the observed source vertex-count set:

\[
\{29,31,34,36,38,40,52\}.
\]

The supplementary table reports 124 source samples, 2.73× volume compression, eight source change points, and 2.99 s solve time. The transition-alignment audit reports a maximum transition error of 0 frames on the tested CS2 captures --- every detected topology transition coincides exactly with a source transition --- so precision and recall are both 1.0, independent of the frame tolerance \(\delta\) (cf.~§2).

\begin{figure*}[t]
\centering
\includegraphics[width=0.95\textwidth]{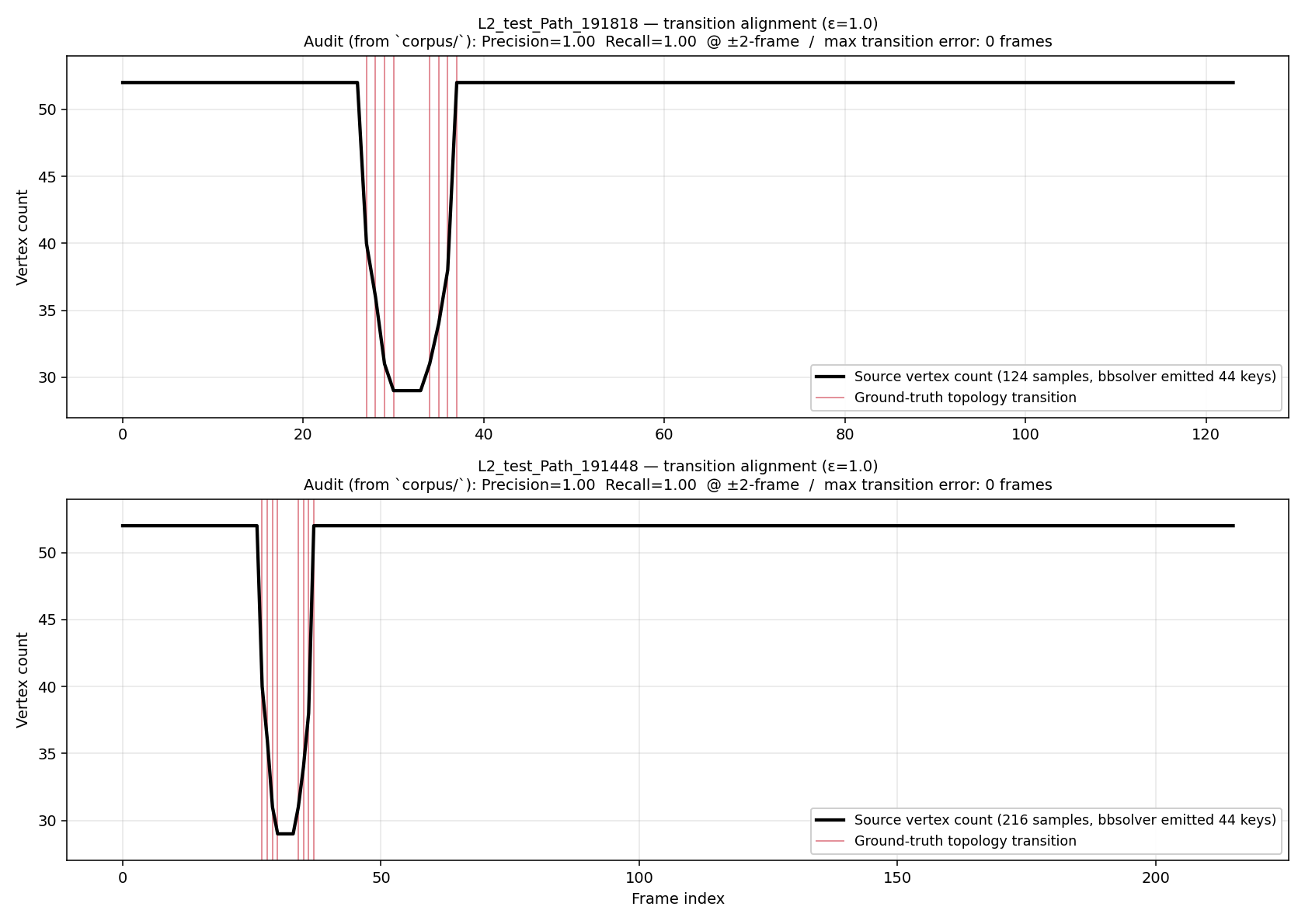}
\caption{CS2 topology transition alignment.}\label{fig:cs2}
\end{figure*}

This is not presented as the AE production writeback mode. It demonstrates that the solver can detect and track topology regions in a variable-topology source trajectory (Fig.~\ref{fig:cs2}).

\subsection{Procedural noodle path and blob lineage}\label{procedural-noodle-path-and-blob-lineage}

The FK noodle fixture is a production-style variable-topology path expression whose production solve writes AE-compatible uniform-topology output. Path bundles often show a small gap between the solver-side and AE-side verify, because AE densifies path evaluation differently than the solver's polyline projection.

At \(\varepsilon=1\), the two relevant noodle outputs are:

\begin{itemize}
\tightlist
\item
  \textbf{Production two-pass:} 83 keys from 242 samples, 6.7× volume reduction, 22 output vertices, \texttt{solver\_max\_err\ =\ 0.974} px / \texttt{cli\_verify\_max\_err\ =\ 0.974} px / \texttt{ae\_roundtrip\_max\_err\ =\ 1.160} px, 13.0 s solve time.
\item
  \textbf{Experimental per-region mode:} 139 main-path keys, 4.4× volume reduction, 20 output vertices, \texttt{solver\_max\_err\ =\ 0.982} px / \texttt{cli\_verify\_max\_err\ =\ 1.204} px (one over-budget landmark-subpath emission) / \texttt{ae\_roundtrip\_max\_err\ =\ 3.990} px, 394.0 s solve time. (This 139 counts the production-relevant main path only; the experimental flag set also emits a diagnostic \texttt{landmark\_subpath} property, so \texttt{corpus\_manifest.csv} reports a bundle total of 276 keys = 139 main + 137 landmark.)
\end{itemize}

\begin{figure*}[t]
\centering
\includegraphics[width=0.95\textwidth]{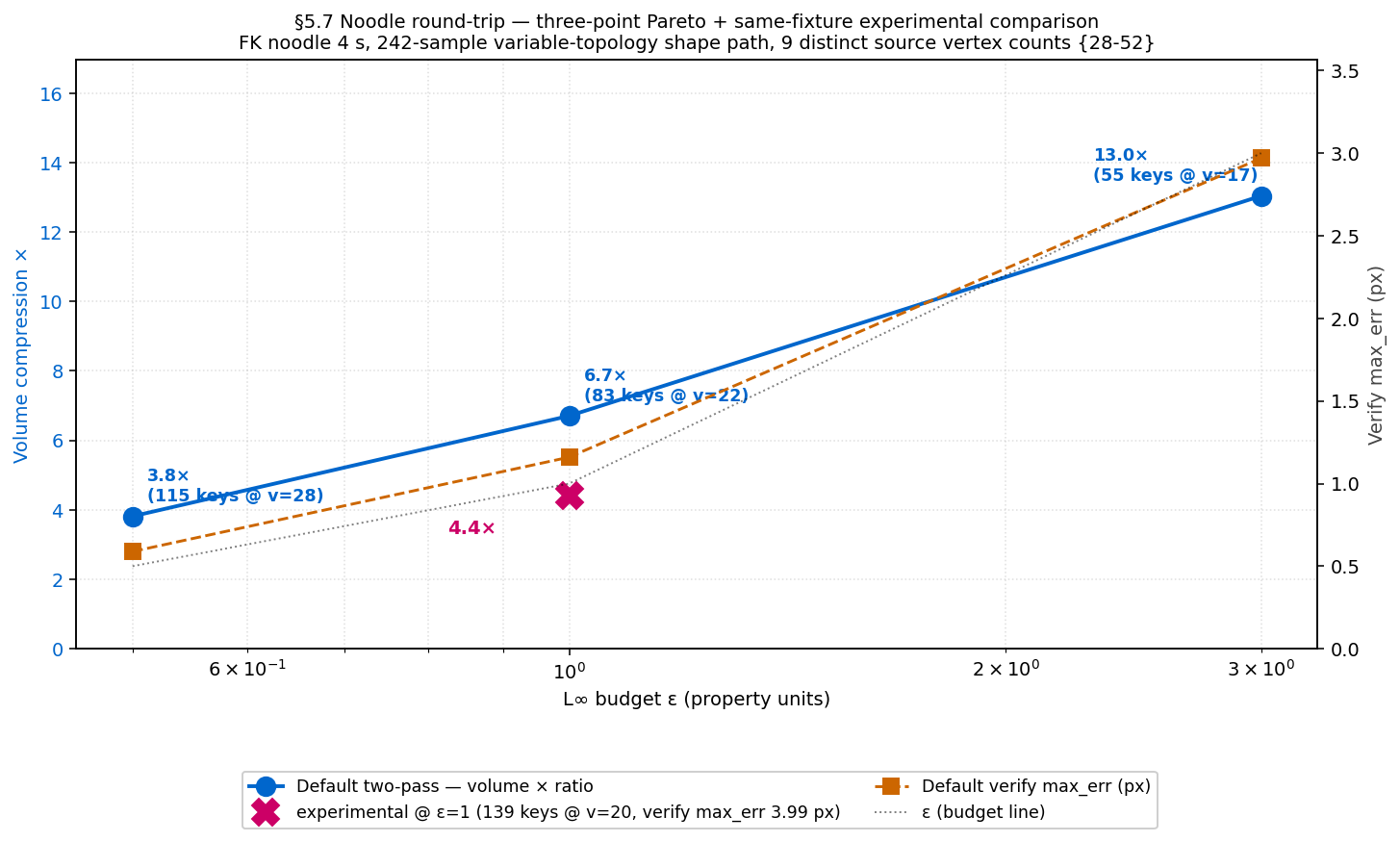}
\caption{Noodle tolerance sweep.}\label{fig:noodle-pareto}
\end{figure*}

\begin{figure*}[t]
\centering
\includegraphics[width=0.95\textwidth]{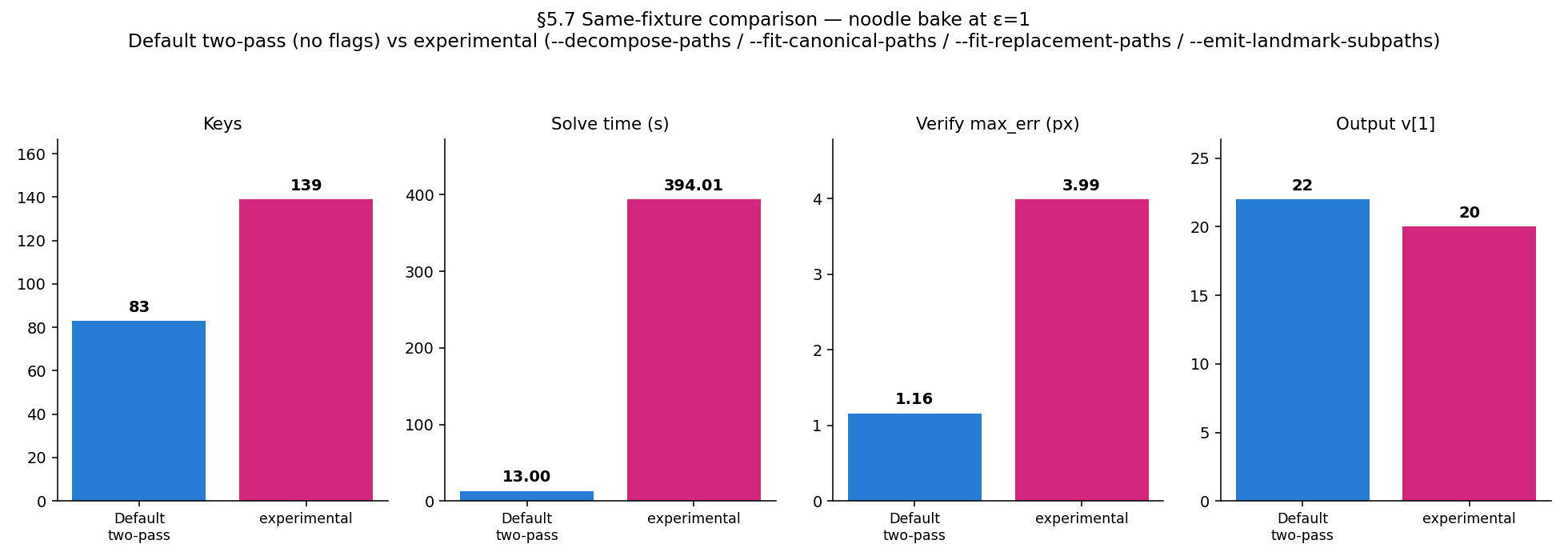}
\caption{Noodle production two-pass vs experimental per-region mode.}\label{fig:noodle-bars}
\end{figure*}

The experimental mode is included as a negative result: smaller per-key topology did not make the overall output better. It used more keys, ran about 30× longer (394.0 s vs 13.0 s), and verified worse in AE (3.990 px vs 1.160 px). The better future direction is to keep the production two-pass architecture and add optional per-region vertex refinement only when it improves the validated cost. Fig.~\ref{fig:noodle-pareto} and Fig.~\ref{fig:noodle-bars} contrast the production and experimental modes.

\begin{figure*}[t]
\centering
\includegraphics[width=0.95\textwidth]{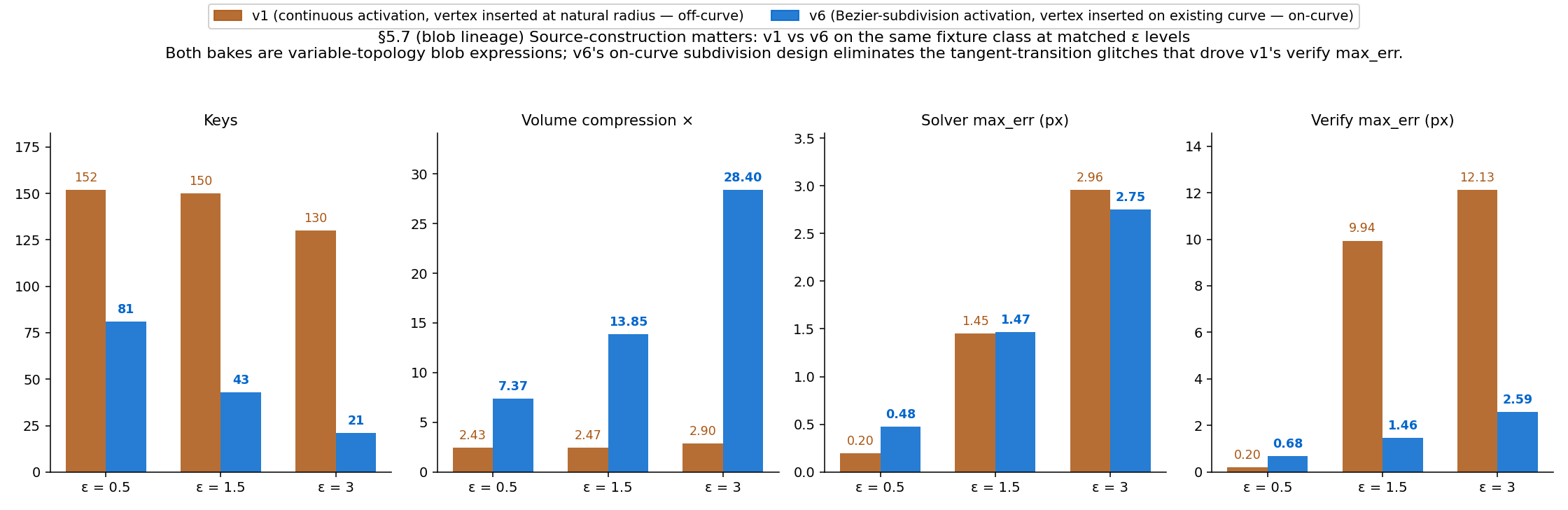}
\caption{Variable-topology blob lineage comparison.}\label{fig:blob}
\end{figure*}

A longer blob expression illustrates source-design sensitivity. A continuous off-curve activation creates many topology transitions and large AE round-trip gaps at loose budgets, whereas a subdivision-based on-curve formulation produces cleaner topology changes and tighter round-trip behavior. At \(\varepsilon=1.5\), the v6 subdivision version emits 43 keys, reaches 13.85× volume compression, with \texttt{ae\_roundtrip\_max\_err\ =\ 1.465} px (Fig.~\ref{fig:blob}).

\subsection{Synthetic topology fixtures}\label{synthetic-topology-fixtures}

The repository also contains synthetic topology-event visuals. These are useful for explaining why per-frame simplifiers can pop --- abruptly change vertex count between adjacent frames --- or introduce spurious topology states. Because its source summary JSON is not part of the public supplementary manifest alongside the published CSVs, this paper treats the synthetic figure as qualitative. The main quantitative variable-topology claims in this paper rely on CS2, noodle, and blob artifacts whose tables are present in the public supplementary data.

\subsection{Determinism}\label{determinism}

The audit reports 60 invocations: 3 fixtures × 2 schedulers (serial, multi-threaded) × 10 repetitions. Each (fixture, scheduler) condition produced exactly one unique \emph{normalized} output hash --- all 10 repetitions emitted identical \texttt{bbky.json} content after normalization. Normalization removes the run-varying provenance fields (\texttt{request\_id}, \texttt{solve\_time\_ms}, \texttt{solver\_build}, and timestamps), recursively sorts object keys, and hashes the canonical serialization; two runs are identical iff their normalized hashes match. The audit covers \texttt{walk\_cycle\ Rotation} (axis B, ε=0.5), \texttt{bouncing\_ball\_2d} (axis C, ε=1.0), and \texttt{L2\_test\_Path\ stable\ shape} (axis A, ε=1.0). The aggregate ships at \texttt{supplementary/determinism\_audit.csv} and is regenerable from its recorded source \texttt{supplementary/determinism.json} via \texttt{scripts/generate\_determinism\_audit.py}. These identical hashes support the reproducibility claim for the cited bundle outputs, subject to normal hardware-dependent timing differences.

\section{Related work}\label{related-work}

Prior art spans several literatures, so a one-sided framing under-cites: treating the project as only curve simplification omits animation key reduction, and treating it as only key reduction omits editable vector and roto systems.

\subsection{Static curve and path simplification}\label{static-curve-and-path-simplification}

Ramer \hyperref[ref-1]{{[}1{]}} and Douglas-Peucker \hyperref[ref-2]{{[}2{]}} are foundational polygonal approximation methods. Schneider \hyperref[ref-3]{{[}3{]}} is the standard reference for automatically fitting digitized curves with Bezier segments. Wang et al.~\hyperref[ref-15]{{[}15{]}} provide a modern Bezier-spline simplification method using locally integrated error metrics. These works underpin the path side of bbsolver, but they are primarily spatial simplification methods; they do not solve host-evaluated animation streams or produce round-trip-validated sparse bakes.

\subsection{Dense-to-sparse animation keyframe reduction}\label{dense-to-sparse-animation-keyframe-reduction}

Roberts et al.~\hyperref[ref-4]{{[}4{]}} is the most important fixed-topology research neighbor: it formulates optimal and interactive keyframe selection for motion capture and compares against commercial simplification. Earlier work by Xiao et al.~\hyperref[ref-5]{{[}5{]}}, Lee et al.~\hyperref[ref-6]{{[}6{]}}, and Miura et al.~\hyperref[ref-7]{{[}7{]}} belongs to the same family. The overlap is clear: dense motion data is reduced to editable keyframes. The difference is that bbsolver is organized around host-facing property streams, JSON exchange, round-trip host evaluation, and vector path support rather than skeletal/mocap-only data.

\emph{No experimental comparison against Roberts et al.~is included in this paper.} Their reference implementation (the \emph{Salient Poses} tools) is openly available under an MIT license, but only as a Maya-integrated workflow, not a standalone-portable, algorithm-only package. A host-agnostic reproduction would therefore require reimplementing the optimal/interactive key-selection algorithm against our SampleBundle/KeyBundle boundary. We instead tested the three open-source baselines whose algorithm cores are already standalone-portable (Blender Decimate, the Joosten reducer, and the Toolchefs \texttt{keyReducer}). The full §5.3 comparison runs from a clone: the raw FBX \texttt{SampleBundle} ships under \texttt{fbx\_mocap\_retarget\_full\_size/pose\_sampled\_blender\_action/} and is auto-resolved by \texttt{scripts/\_paths.py}, so the bbsolver, Joosten, and Toolchefs branches re-solve directly from the shipped input; the per-tolerance sweep outputs and CSVs are also shipped for direct verification without re-solving. The only part of §5.3 that requires an external install is the Blender F-Curve Decimate baseline, whose \texttt{bpy.ops.graph.decimate} sweep needs a Blender 4.5 install (its outputs ship under \texttt{sweep\_outputs/}). See §5.3 and the fixture \texttt{README}. A from-paper port of Roberts et al.~is plausible follow-up work and is listed in §7 alongside the Wang et al.~spline-simplification port.

\subsection{Host-native and open-source reducers}\label{host-native-and-open-source-reducers}

Blender F-Curve Decimate \hyperref[ref-20]{{[}20{]}}, Maya Simplify Curve / Key Reducer \hyperref[ref-18]{{[}18{]}}, MotionBuilder Key Reducing \hyperref[ref-19]{{[}19{]}}, Toolchefs \texttt{keyReducer} \hyperref[ref-21]{{[}21{]}}, and Joosten's Maya keyframe reducer \hyperref[ref-22]{{[}22{]}} are the closest practical tool neighbors. They overlap on baked-curve cleanup and editability. The FBX comparison shows that local simplification criteria can violate a global sample-space budget: every tested baseline exceeded \(\varepsilon\) on at least one property. That result should be read as a property of the tested fixture and implementation ports, not as a claim that all host-native reducers are inferior in every workflow.

\subsection{Editable spline and roto systems}\label{editable-spline-and-roto-systems}

Agarwala et al.~\hyperref[ref-10]{{[}10{]}}, Roto++ \hyperref[ref-11]{{[}11{]}}, Bermudez et al.~\hyperref[ref-12]{{[}12{]}}, Joint Stroke Tracing and Correspondence \hyperref[ref-13]{{[}13{]}}, and RotoShop \hyperref[ref-14]{{[}14{]}} are important neighbors for editable vector output and temporal correspondence. These systems often start from video, raster masks, or drawings and produce editable splines or strokes. bbsolver starts from arbitrary host-sampled numerical streams and optimizes for sparse host-key output. The concerns overlap, but the input and validation boundary differ.

\subsection{Vector animation with topology changes}\label{vector-animation-with-topology-changes}

The Vector Animation Complex (VAC) \hyperref[ref-8]{{[}8{]}} is the key reference for vector graphics animation with time-varying topology. Tutrace \hyperref[ref-9]{{[}9{]}} is a close recent system for editable animated vector shapes from video. VAC and Tutrace combine vector geometry and time in a broad sense, but they target topology-aware vector animation representations or video-to-vector conversion rather than a host-sampled dense-to-sparse key solve. bbsolver does not supersede them. Its narrower integration problem is reducing sampled host/path streams to sparse editable key representations under one explicit validation budget and host-writeback constraint.

\subsection{Sparse animation representations and interchange}\label{sparse-animation-representations-and-interchange}

OpenUSD spline animation material \hyperref[ref-16]{{[}16{]}} and NVIDIA Omniverse's TimeSample-to-Animation-Curve tool \hyperref[ref-17]{{[}17{]}} are relevant because they formalize or operationalize conversion between dense time samples and sparse animation curves. They are not direct bbsolver equivalents: neither publicly offers topology-aware vector-path solving, and neither reports host round-trip evaluation of the kind shown here for After Effects.

\section{Discussion and limitations}\label{discussion-and-limitations}

\textbf{Systems contribution, not new math.} The paper is strongest when it is evaluated as a production-facing sparse-conversion system. The mathematical notation defines the problem and acceptance checks; it is not intended to disguise established algorithms as new theory.

\textbf{Discrete validation.} The conformance proposition applies to the validation samples. Continuous-time error between samples depends on sampling density, host interpolation, and motion frequency. Sub-frame sampling is supported by the schema but needs broader evaluation.

\textbf{Host fidelity.} AE round-trip fidelity is currently the strongest host evidence. The Blender/FBX path adds cross-host sampling and transform-curve evidence, but full Bezier tangent round-trip through stock Blender FBX I/O is limited by Blender's importer/exporter behavior. Maya, MotionBuilder, or SDK-level FBX validation would be useful future work.

\textbf{Variable topology.} Variable-topology path streams are less common in ordinary hand-keyed AE path workflows because manual vertex additions/removals usually change the whole path's vertex set rather than producing per-key vertex birth/death. They are still relevant for expressions, procedural generators, auto-trace/vectorization, drawing-substitution workflows, and geometry/cache imports. The paper treats variable-topology cases as stress tests and procedural-source evidence, not as the most common animator interaction.

\textbf{Baselines.} The baseline landscape is broad. The current public comparison covers three open-source reducers on one FBX fixture and Illustrator on static SVG fixtures. Commercial Maya/MotionBuilder/Omniverse comparisons would strengthen the fixed-topology side, and video-to-vector systems such as Tutrace/RotoShop would strengthen the vector-shape discussion.

\textbf{Performance.} bbsolver is accuracy-first and offline. Worst-case solves can be expensive, especially path-heavy or experimental topology modes. This is acceptable for offline baking workflows but rules out positioning bbsolver as an interactive simplifier.

\subsection{Limitations and considerations}\label{limitations-and-considerations}

This section gives a short, explicit list of where the evidence could be stronger or misread, so a reader can weigh the claims against their own context.

\begin{itemize}
\tightlist
\item
  \textbf{Fixture selection.} The headline rigs (DUIK humanoid, ant hexapod, FBX mocap retarget) were chosen for diversity of property type and source provenance, not by random sampling from a registered fixture pool. They are representative of the production problem class but not a statistical sample.
\item
  \textbf{Baseline algorithmic representation.} The Joosten reducer and Toolchefs \texttt{keyReducer} are standalone-Python ports of Maya plugins; the ports preserve the algorithmic core and ship original source for diff verification, but a small chance remains that subtle Maya-specific behavior (e.g.~host pre-filtering of source curves) would change results inside Maya. The Blender Decimate comparison uses Blender's stock \texttt{bpy.ops.graph.decimate}, run headlessly.
\item
  \textbf{Key-count metric.} The level-playing-field key counts are reported both as animator edit points (where bbsolver's shared-timing ThreeD keys count once) and as scalar-equivalent storage (each bbsolver vector key counted ×3). Both numbers appear in §5.3; readers should not quote the vector-metric ratio as a raw storage advantage.
\item
  \textbf{Three distinct verification notions.} The supplementary CSVs and corpus split the achieved error into three columns wherever the three values differ: \texttt{solver\_max\_err} (in-loop validation against sampled \texttt{bbsm} values during the solve), \texttt{cli\_verify\_max\_err} (regenerated by \texttt{bbsolver\ verify\ \textless{}bbky\textgreater{}\ \textless{}bbsm\textgreater{}} with v1.0.1; recorded in each \texttt{\textless{}req\textgreater{}.verify.json}), and \texttt{ae\_roundtrip\_max\_err} (After Effects writeback then re-sampled at the same source times). For transform rigs these three numbers agree to within rounding; for \texttt{shape\_flat} path rows the AE round-trip can be looser than the solver-internal verify because AE's path evaluator densifies differently than the solver's polyline projection (e.g.~noodle ε=1 reports \texttt{solver\_max\_err\ =\ 0.974} and \texttt{ae\_roundtrip\_max\_err\ =\ 1.160}). Quote the column that matches your reproduction path.
\item
  \textbf{Path Hausdorff sampling density.} The contour error metric for \texttt{shape\_flat} paths is computed on a densely-sampled polyline approximation, not analytic Bezier-to-Bezier distance. The dense sampler is configured at a density that empirically converges within a small fraction of ε in our tests, but we did not tabulate a full sensitivity sweep over sampling density.
\item
  \textbf{``Host-agnostic'' empirical scope.} The host contract (JSON SampleBundle in / KeyBundle out) is host-agnostic by design. The published empirical validation covers two real hosts (After Effects, Blender) plus a host-less FBX direct path; further host integrations (Maya, MotionBuilder, Houdini, Nuke) are future work and not validated in this paper.
\item
  \textbf{Proprietary tooling in evidence path.} Reproducing the AE round-trip section requires a valid Adobe After Effects 2024+ license; reproducing the Illustrator SVG comparison requires an Illustrator install. The bbsm bundles in \texttt{corpus/} provide the AE-independent path so solver-only reproduction is fully open. Section 5.2 results that depend on AE re-evaluation cannot be reproduced without the AE license.
\end{itemize}

\clearpage\onecolumn\section{Reproducibility and artifacts}\label{reproducibility-and-artifacts}

The public repository \hyperref[ref-23]{{[}23{]}} is:

\begin{quote}
https://github.com/ivg-design/bbsolver
\end{quote}

The bbsolver source repository is MIT licensed; this companion artifact is mixed-license, combining MIT-licensed IVG Design scripts and adapters, third-party baseline code under its own licenses, and manuscript/data artifacts under the terms stated in \texttt{LICENSE} and \texttt{THIRD\_PARTY\_NOTICES.md}. The artifact is self-contained for paper inspection, shipped-corpus verification, figure/table audit, and solver-only reproduction of the redistributed bundles; reproducing the host round-trips additionally requires the external \texttt{bbsolver} binary, After Effects, Blender, or Illustrator as noted per section. The manuscript ships alongside all of its figures, supplementary CSVs, the raw solve corpus, baseline runners, deterministic scripts, and host adapters, organized as follows (Table~\ref{tbl:artifacts}):

\begingroup\scriptsize

\begin{longtable}[]{@{}
  >{\raggedright\arraybackslash}p{(\columnwidth - 4\tabcolsep) * \real{0.2600}}
  >{\raggedright\arraybackslash}p{(\columnwidth - 4\tabcolsep) * \real{0.4000}}
  >{\raggedright\arraybackslash}p{(\columnwidth - 4\tabcolsep) * \real{0.3400}}@{}}
\caption{Companion artifact archive layout. All paths are relative to this artifact's root.}\label{tbl:artifacts}\tabularnewline
\toprule\noalign{}
\begin{minipage}[b]{\linewidth}\raggedright
Path
\end{minipage} & \begin{minipage}[b]{\linewidth}\raggedright
Contents
\end{minipage} & \begin{minipage}[b]{\linewidth}\raggedright
Reproducibility role
\end{minipage} \\
\midrule\noalign{}
\endfirsthead
\toprule\noalign{}
\begin{minipage}[b]{\linewidth}\raggedright
Path
\end{minipage} & \begin{minipage}[b]{\linewidth}\raggedright
Contents
\end{minipage} & \begin{minipage}[b]{\linewidth}\raggedright
Reproducibility role
\end{minipage} \\
\midrule\noalign{}
\endhead
\bottomrule\noalign{}
\endlastfoot
\texttt{supplementary/manifest.csv} & Index of every supplementary CSV with its paper section and source request\_ids. & Entry point --- open this file first. \\
\texttt{corpus/} + \texttt{corpus\_manifest.csv} & Raw \texttt{bbsm} / \texttt{bbky} / \texttt{verify} / \texttt{progress.log} bundles for the 11 cited solve requests --- 16 \texttt{bbky} files in total, since some solves emit multiple property groups (\textasciitilde62 MB). & Solver re-run reproduces the published key counts and residuals exactly for the same build. \\
\texttt{corpus/THRESHOLD\_NOTE.md} & Documents which \texttt{bbsolver} version is required to reproduce the canonical \texttt{verify.json} (v1.0.1) and why v1.0.0's verifier had an over-strict per-dimension bug on variable-topology \texttt{shape\_flat}. & Read this before downloading binaries. \\
\texttt{supplementary/production\_corpus\_per\_run.csv} + \texttt{production\_corpus\_summary.csv} & Per-run + aggregate stats for the 203-run §5.1 corpus. & Backs §5.1 reduction/solve-time claims. \\
\texttt{supplementary/table\_*.csv}, \texttt{per\_property\_*.csv}, \texttt{fbx\_mocap\_*.csv}, \texttt{svg\_decimation\_*.csv} & One CSV per quantitative table / figure. & Direct verification without re-running the solver. \\
\texttt{supplementary/determinism.json} + \texttt{determinism\_audit.csv} & §5.9 determinism source data and aggregate audit. & Regenerable via \texttt{scripts/}; backs §5.9. \\
\texttt{external\_runners/joosten\_reducer/} + \texttt{toolchefs\_reducer/} & Standalone-Python ports of the two Maya-plugin baselines, with upstream source preserved for diff. & Open-source comparison runs without a Maya license. \\
\texttt{fbx\_mocap\_retarget\_full\_size/pose\_sampled\_blender\_action/retarget\_full\_size.bbsm.json} & Raw input \texttt{SampleBundle} for the §5.3 cross-host comparison (45 properties, 13,455 source samples; Mixamo-derived retarget), auto-resolved by \texttt{scripts/\_paths.py}. & Re-solve the bbsolver / Joosten / Toolchefs §5.3 branches end to end from a clone. \\
\texttt{fbx\_mocap\_retarget\_full\_size/sweep\_outputs/} & Per-method, per-tolerance §5.3 solver outputs for all four methods. & Re-derive the §5.3 comparison tables and figures, or verify directly without re-solving. \\
\texttt{scripts/} & Deterministic Python pipeline (figure / table generators, \texttt{smoke\_reproduce\_one\_row.py}, \texttt{generate\_determinism\_audit.py}, and a \texttt{\_paths.py} portability helper). & Regenerate the corpus-derived figures and CSVs from the raw bundles. The \texttt{ae\_roundtrip\_max\_err} columns and the §5.2 / §5.7 round-trip figures embed After Effects host-playback measurements that are not recomputable from the CLI corpus, and the 203-run §5.1 aggregate plus its \texttt{production\_compression\_vs\_tolerance.png} boxplot derive from private development runs (the per-run requested \(\varepsilon\) is not exported to the public CSV); the generators preserve these shipped values rather than recomputing them (see their headers). \\
\texttt{fixtures/blob\_variable\_topology\_*.js} & v1--v6 lineage of AE Property Expressions for §5.7 blob tests. & Source-construction evidence. \\
\texttt{fixtures/svg\_decimation/} & 8 generated SVGs + Illustrator outputs + \texttt{generate\_test\_svgs.py}. & Reproducible inputs for §5.4. \\
\texttt{examples/after-effects/bbsolver-test-harness.jsx} + \texttt{bbsolver-test-harness/} modules & The AE ScriptUI reference adapter cited in §3.1 (main script plus its included \texttt{.jsx} modules); Blender adapter scripts under \texttt{examples/blender/}. & Source for the host round-trip harness. \\
\texttt{docs/AE\_ROUND\_TRIP\_CAPTURE\_GUIDE.md} & Operator protocol for capturing AE round-trip artifacts via the public harness. & Replicate the §5.2 round-trip on additional fixtures. \\
\texttt{docs/NOODLE\_EXPRESSION\_ANALYSIS.md} & Static analysis of the noodle expression's topology states. & Source-level grounding for §5.7. \\
\texttt{after\_effects\_benchmark\_project/} --- \texttt{bbSolver\_benchmarking.aep} & After Effects project with the compositions for every §5.2 / §5.5 / §5.6 / §5.7 fixture. & Replicate the full host round-trip (requires Adobe AE 2024+ license). \\
\texttt{LICENSE} + \texttt{THIRD\_PARTY\_NOTICES.md} & MIT license for the IVG Design-authored scripts/adapters; per-component terms for the bundled reducers (Joosten: MIT; Toolchefs: LGPL-3.0) and the manuscript/corpus. & Licensing provenance for redistribution. \\
\end{longtable}

\endgroup

The solver build used to produce the keys in the published corpus is \texttt{bbsolver\ 1.0.0}, tag \texttt{v1.0.0} (commit \texttt{5e29545}). Each \texttt{bbky.json} records \texttt{solver\_version} (\texttt{bbsolver\ 1.0.0}) and \texttt{solver\_build} (the literal string \texttt{dev}); the \texttt{solver\_build} field does not encode the tag/commit, so the exact provenance is the tag/commit stated here and is checkable by re-solving with the tagged source rather than read from the bundle. (One experimental §5.7 bundle, \texttt{req-1779741201109\_g1}, was produced by a development snapshot and records \texttt{bbsolver\ 0.1.0}; all 15 other corpus bundles record \texttt{bbsolver\ 1.0.0}.) The canonical \texttt{verify.json} files were regenerated with \texttt{bbsolver\ 1.0.1} (tag \texttt{v1.0.1}, commit \texttt{7342c0d}), which fixes an overly strict per-dimension check in the verifier that had wrongly rejected the six variable-topology \texttt{shape\_flat} rows of the corpus (noodle and blob fixtures). v1.0.1 is a verifier bug fix, not a solver numeric change: for the 15 \texttt{bbsolver\ 1.0.0} corpus bundles, re-solving with either tag against the same \texttt{bbsm.json} reproduces the same key counts and \texttt{max\_err} to six decimal places. Raw \texttt{bbky.json} bytes still differ in the run-varying provenance/timing fields (\texttt{request\_id}, \texttt{solve\_time\_ms}, \texttt{solver\_build}, timestamps), so reproduction is checked on the §5.9 normalized output, not a raw \texttt{diff} (the experimental \texttt{0.1.0} row is verified, not claimed byte-reproducible from a v1.0.0 build). See \texttt{corpus/THRESHOLD\_NOTE.md} for the full diagnosis. Timings are hardware-dependent; key counts and residuals are deterministic for the same solver build and input bundle. To rebuild from source, \texttt{git\ checkout\ v1.0.0} (or \texttt{v1.0.1}) rather than a raw commit hash.

\section{Disclosures}\label{disclosures}

\textbf{Generative AI.} This project was developed through AI-augmented software engineering by an independent motion-design tool developer. Generative AI tools assisted with C++ implementation scaffolding, debugging suggestions, documentation drafting, research planning, and manuscript organization. The human author originated the production problem, specified solver behavior, reviewed and modified all code, ran and interpreted the benchmarks, and verified every citation and claim. The author assumes full responsibility for the released code and manuscript. No AI system is an author. The paper presents no AI-generated or fabricated measurements, data, or citations.

\textbf{Note on the name.} ``bbsolver'' is unrelated to the several block-breaker (``BB'') puzzle solvers that share the name; here ``bb'' is short for bakerBoy, the project bbsolver grew out of.

\textbf{Competing interests.} The author is developing bakerBoy, a commercial After Effects extension built around bbsolver. The solver itself is released independently under the MIT License; this paper reports only on the open-source solver.

\section{Conclusion}\label{conclusion}

bbsolver is best understood as a sparse-conversion system organized around a host-agnostic JSON contract, not as a new mathematical simplification primitive. The empirical evidence in this paper validates that contract against two real hosts, After Effects and Blender, plus a host-less FBX path. Additional host integrations remain future work. Its strongest contribution is tolerance-bounded conversion of fixed-topology streams into sparse editable keys, validated by AE round-trip rigs and the cross-host Blender/FBX comparison. Its path contribution extends the same contract to vector paths, including constant-topology path reduction and topology-aware diagnostics for harder procedural inputs.

The result is a reproducible companion artifact, built around an MIT-licensed open-source solver, for a common production problem: converting dense evaluated animation into editable host-native keys within a measured accuracy budget.

\clearpage
\onecolumn
\appendix
\section*{Appendix overview}
The following sections contain operator-level reproduction details that are useful for readers who want to run the artifact, but are not needed for the main argument of the paper.

\section{Pre-built release binaries}\label{appendix-pre-built-release-binaries}

The \texttt{v1.0.1} GitHub release ships pre-built \texttt{bbsolver} binaries for four (host, architecture) combinations. Readers who do not want to compile from source can use them to reproduce the canonical CLI verify pass on the full paper corpus. All four v1.0.1 artifacts share the same source tree at tag \texttt{v1.0.1}, and a single SHA-256 manifest covers the set. The \texttt{v1.0.0} release remains available alongside (with the same four platforms) for byte-level reproduction of the solver outputs themselves, but its \texttt{bbsolver\ verify} CLI cannot canonical-verify the six variable-topology paper-corpus rows; use v1.0.1 to reproduce \texttt{verify.json} from scratch.

All four (host, architecture) combinations are built from the same source commit per release; both releases target macOS 15+ (Apple Clang, Xcode 16) for the macOS tarballs and Windows 11 (MSVC 2022) for the Windows zips. The Windows x64 binary is cross-compiled from an arm64 host via \texttt{vcvarsall.bat\ arm64\_amd64}. Sizes on disk: the macOS tarballs are \textasciitilde14.8 MB (arm64) and \textasciitilde15.0 MB (x86\_64); the Windows zips are \textasciitilde22.8 MB (arm64) and \textasciitilde23.4 MB (x64).

The SHA-256 manifests below are formatted to be directly consumable by \texttt{shasum\ -a\ 256\ -c} and \texttt{(Get-FileHash\ -Algorithm\ SHA256)}. Hashes are integrity anchors only; reproducible-from-source rebuilds + the self-checking paper corpus (see Integrity vs authenticity below) provide the authenticity layer.

\Needspace*{9\baselineskip}

\textbf{v1.0.1 --- canonical for \texttt{bbsolver\ verify}:}

\begin{Verbatim}[fontsize=\scriptsize,frame=single,framesep=2mm,xleftmargin=0pt]
c28ce4c9530f298b094a7ec35305ca39dc35d25f9adf2014c7f30cb69deb26ad  bbsolver-v1.0.1-macos-arm64.tar.gz
6e221df339dd89366ad38ca3a29dbf4501550d3be73d9bb863cd2eb6f3167d5e  bbsolver-v1.0.1-macos-x86_64.tar.gz
776f0f8278aa76fd0fdcb3f45bc0ce1d62d68ba8eb31b833f8ced1964c067f08  bbsolver-v1.0.1-windows-arm64.zip
daa6448a888b18820b18971d4465c4c1668414bc2f75ad6f220a458bc7d5e24c  bbsolver-v1.0.1-windows-x64.zip
\end{Verbatim}

\Needspace*{9\baselineskip}

\textbf{v1.0.0 --- solver of record for the shipped \texttt{bbky.json} corpus:}

\begin{Verbatim}[fontsize=\scriptsize,frame=single,framesep=2mm,xleftmargin=0pt]
24849db7d7ac3e74d147b7ff4f154558671125fe305051d503b3b2fe40f69b7d  bbsolver-v1.0.0-macos-arm64.tar.gz
45e42d5a5eec6df3cf0bde890efbcc0ae8e1ab9469dab577ce746be974d2c588  bbsolver-v1.0.0-macos-x86_64.tar.gz
bb22308bc500f6a718e3faf99718cd87cec07dbd8683ecf510d8706a648992f7  bbsolver-v1.0.0-windows-arm64.zip
938809d6036a4dbc4caa6f0627db4dc1cf32ee53729a5b12d9aacffdc81e10f1  bbsolver-v1.0.0-windows-x64.zip
\end{Verbatim}

\textbf{Download + verify (macOS / Linux):}

\begin{verbatim}
gh release download v1.0.1 --repo ivg-design/bbsolver \
  --pattern 'bbsolver-v1.0.1-macos-arm64.tar.gz' \
  --pattern 'SHA256SUMS.txt'
shasum -a 256 -c SHA256SUMS.txt --ignore-missing
# Expected: bbsolver-v1.0.1-macos-arm64.tar.gz: OK
tar -xzf bbsolver-v1.0.1-macos-arm64.tar.gz
./bbsolver-v1.0.1-macos-arm64/bin/bbsolver --version
# bbsolver 1.0.1
\end{verbatim}

\textbf{Download + verify (Windows PowerShell):}

\begin{verbatim}
gh release download v1.0.1 --repo ivg-design/bbsolver `
  --pattern 'bbsolver-v1.0.1-windows-x64.zip' `
  --pattern 'SHA256SUMS.txt'
(Get-FileHash -Algorithm SHA256 .\bbsolver-v1.0.1-windows-x64.zip).Hash.ToLower()
# Compare against the matching line in SHA256SUMS.txt
Expand-Archive .\bbsolver-v1.0.1-windows-x64.zip
.\bbsolver-v1.0.1-windows-x64\bin\bbsolver.exe --version
\end{verbatim}

\textbf{Integrity vs authenticity.} SHA-256 sums establish \emph{integrity} (the bytes you downloaded match the bytes the release uploaded) but not \emph{authorial authenticity}: a matching hash does not by itself prove the release came from the author, because anyone who could replace the release artifacts could also replace the published hashes. We rely on three layers of trust:

\begin{enumerate}
\def\labelenumi{\arabic{enumi}.}
\tightlist
\item
  \textbf{GitHub release transport.} The release page is served over HTTPS; the CDN authenticates the connection with GitHub's TLS certificate at delivery time. This is the same guarantee that protects \texttt{gh\ release\ download}.
\item
  \textbf{Reproducible-from-source.} All four binaries are built from the public source at tag \texttt{v1.0.0} (commit \texttt{5e29545}); the v1.0.1 binaries are built from tag \texttt{v1.0.1} (commit \texttt{7342c0d}). Anyone can \texttt{git\ checkout} the tag, rebuild, and recompute the SHA-256 of their local build. Compiler-determinism caveats apply (build path, timestamps, parallel link order), so a bit-for-bit reproduction requires the same toolchain version; the recorded toolchain versions in the table above are deliberately specific to allow that comparison.
\item
  \textbf{The corpus is self-checking.} Every \texttt{bbky.json} in \texttt{corpus/} records its \texttt{solver\_version} and \texttt{solver\_build}. Running any released binary against the matching \texttt{bbsm.json} input must produce the same \texttt{total\_keys} and per-property \texttt{max\_err} recorded in \texttt{verify.json}; the hash comparison of normalized output is exact.
\end{enumerate}

\textbf{Not signed.} These binaries are \textbf{not} code-signed for distribution: there is no Apple Developer ID notarization on the macOS tarballs and no Authenticode / EV certificate on the Windows zips. macOS Gatekeeper, Windows SmartScreen, and AV heuristics on either platform may warn on first run, and the operator must either explicitly allow the binary, build from source, or skip the pre-built path and use the supplementary CSVs as the alternative reproducibility anchor. Code signing is on the future-work list (see §7 limitations).

\textbf{Toolchain notes from local builds.} The v1.0.1 macOS binaries were built natively on Apple Silicon with the macOS SDK; the x86\_64 macOS binary uses \texttt{CMAKE\_OSX\_ARCHITECTURES=x86\_64} cross-compile. Both ran the full v1.0.1 ctest suite, 117/117 passing, including the new variable-topology \texttt{shape\_flat} verifier tests and the paper-corpus integration test that exercises canonical CLI verify against all 16 \texttt{bbky} bundles in \texttt{corpus/}. The Windows binaries were both built inside a Windows 11 ARM64 VMware Fusion VM: the arm64 binary natively, the x64 binary via the \texttt{vcvarsall.bat\ arm64\_amd64} cross-toolchain. All four binaries report \texttt{bbsolver\ 1.0.1} from \texttt{-\/-version}.

\section{One-row reproducibility smoke test}\label{appendix-one-row-reproducibility-smoke-test}

The smallest end-to-end reproduction check ships at \texttt{scripts/smoke\_reproduce\_one\_row.py}. It re-solves a single paper-cited bundle (defaults to the §5.7 noodle ε=1.0 bake at \texttt{corpus/req-1779737483003/}, \textasciitilde5 s wall-clock) and asserts that the produced \texttt{total\_keys} and per-property \texttt{max\_err} match the published bundle within a configurable relative tolerance (default 1 \%). On the author's machine the script reports 0.0000 \% drift. On any platform where bbsolver is built from the same source commit, this is the expected result: the inner DP, fitter, and validator are deterministic for a given solver build and input bundle, so the drift is zero (the 1 \% tolerance only guards against cross-toolchain rounding). The smoke is intentionally lightweight; with a \texttt{bbsolver} binary available (downloaded per §8.1 or built from the public source), run it from this artifact's root:

\begin{verbatim}
# 1. Obtain a bbsolver binary: download a v1.0.1 release (see §8.1), or build it
#    from the public source at https://github.com/ivg-design/bbsolver.
# 2. From this artifact's root, re-solve and verify one published bundle:
python3 scripts/smoke_reproduce_one_row.py --bbsolver /path/to/bbsolver
# expected last line: "PASS: reproduction matches published bundle within tolerance."
\end{verbatim}

A reviewer who only wants to confirm that the published numbers actually came out of the published solver can run this single command and stop.

\small\setlength{\parskip}{0.25em}

\section*{References}\label{references}

\phantomsection\label{ref-1}{}{[}1{]} Urs Ramer. ``An iterative procedure for the polygonal approximation of plane curves.'' \emph{Computer Graphics and Image Processing}, 1972.

\phantomsection\label{ref-2}{}{[}2{]} David Douglas and Thomas Peucker. ``Algorithms for the reduction of the number of points required to represent a digitized line or its caricature.'' \emph{The Canadian Cartographer}, 1973.

\phantomsection\label{ref-3}{}{[}3{]} Philip J. Schneider. ``An Algorithm for Automatically Fitting Digitized Curves.'' In \emph{Graphics Gems}, 1990.

\phantomsection\label{ref-4}{}{[}4{]} Richard Roberts, John P. Lewis, Ken Anjyo, Jaewoo Seo, and Yeongho Seol. ``Optimal and interactive keyframe selection for motion capture.'' \emph{Computational Visual Media}, 2019.

\phantomsection\label{ref-5}{}{[}5{]} Jun Xiao, Yueting Zhuang, Tao Yang, and Fei Wu. ``An Efficient Keyframe Extraction from Motion Capture Data.'' \emph{Computer Graphics International}, 2006.

\phantomsection\label{ref-6}{}{[}6{]} T. Lee, C. Lin, Y. Wang, and T. Chen. ``Animation Key-Frame Extraction and Simplification Using Deformation Analysis.'' \emph{IEEE Transactions on Circuits and Systems for Video Technology}, 2008.

\phantomsection\label{ref-7}{}{[}7{]} T. Miura, T. Kaiga, T. Shibata, H. Katsura, K. Tajima, and H. Tamamoto. ``A Hybrid Approach to Keyframe Extraction from Motion Capture Data Using Curve Simplification and Principal Component Analysis.'' \emph{IEEJ Transactions on Electrical and Electronic Engineering}, 2014.

\phantomsection\label{ref-8}{}{[}8{]} Boris Dalstein, Rémi Ronfard, and Michiel van de Panne. ``Vector Graphics Animation with Time-Varying Topology.'' \emph{ACM SIGGRAPH}, 2015.

\phantomsection\label{ref-9}{}{[}9{]} Loïc Vital. ``Tutrace: Editable Animated Vector Shape from Video.'' \emph{GRAPP/VISIGRAPP}, 2025.

\phantomsection\label{ref-10}{}{[}10{]} Aseem Agarwala, Aaron Hertzmann, David Salesin, and Steven Seitz. ``Keyframe-Based Tracking for Rotoscoping and Animation.'' \emph{ACM Transactions on Graphics}, 2004.

\phantomsection\label{ref-11}{}{[}11{]} Wenbin Li, Fabio Viola, Jonathan Starck, Gabriel J. Brostow, and Neill D. F. Campbell. ``Roto++: Accelerating Professional Rotoscoping using Shape Manifolds.'' \emph{ACM Transactions on Graphics}, 2016.

\phantomsection\label{ref-12}{}{[}12{]} Luis Bermudez, Nadine Dabby, et al.~``A Learning-Based Approach to Parametric Rotoscoping of Multi-Shape Systems.'' \emph{WACV}, 2021.

\phantomsection\label{ref-13}{}{[}13{]} Haoran Mo, Chengying Gao, and Ruomei Wang. ``Joint Stroke Tracing and Correspondence for 2D Animation.'' \emph{ACM Transactions on Graphics}, 2024.

\phantomsection\label{ref-14}{}{[}14{]} S. Ghebremusse. ``RotoShop: Automatic Dataset Rotoscoping via Differentiable Splining.'' \emph{SIGGRAPH Asia Technical Communications}, 2025.

\phantomsection\label{ref-15}{}{[}15{]} S. Wang, C. Liu, D. Panozzo, D. Zorin, and A. Jacobson. ``Bezier Spline Simplification Using Locally Integrated Error Metrics.'' \emph{SIGGRAPH Asia}, 2023.

\phantomsection\label{ref-16}{}{[}16{]} Pixar Animation Studios. ``Spline Animation in USD.'' OpenUSD proposal and development material, 2024-2026.

\phantomsection\label{ref-17}{}{[}17{]} NVIDIA. ``TimeSample to Animation Curve.'' Omniverse documentation.

\phantomsection\label{ref-18}{}{[}18{]} Autodesk. ``Maya Key Reducer Filter'' and ``Simplify Curve'' documentation.

\phantomsection\label{ref-19}{}{[}19{]} Autodesk. ``MotionBuilder Key Reducing Filter'' documentation.

\phantomsection\label{ref-20}{}{[}20{]} Blender Foundation. ``F-Curve Decimate'' documentation.

\phantomsection\label{ref-21}{}{[}21{]} Toolchefs. \texttt{keyReducer} Maya plugin, LGPL open-source repository. https://github.com/toolchefs/keyReducer

\phantomsection\label{ref-22}{}{[}22{]} Robert Joosten. \texttt{maya-keyframe-reduction}, MIT open-source repository. https://github.com/robertjoosten/maya-keyframe-reduction

\phantomsection\label{ref-23}{}{[}23{]} Ilya Gusinski / IVG Design. \texttt{bbsolver} public repository. https://github.com/ivg-design/bbsolver

\end{document}